%% file: thesis.tex
\documentclass[draft,final]{vutinfth} 

\input{preamble.tex}

\newtheorem{theorem}[algocf]{Theorem}
\newtheorem{observation}[algocf]{Observation}
\newtheorem{lemma}[algocf]{Lemma}
\newtheorem{branchingrule}{Branching Rule}
\counterwithout{branchingrule}{section}

\theoremstyle{definition}
\newtheorem{transformation}[algocf]{Transformation}

\begin{document}
\flushbottom
\input{chapters/title}


\subfile{chapters/chapter-introduction.tex}

\subfile{chapters/chapter-preliminaries.tex}

\subfile{chapters/chapter-theoretical-results.tex}

\subfile{chapters/chapter-heuristic-improvements.tex}

\subfile{chapters/chapter-implementation-details.tex}

\subfile{chapters/chapter-experimental-results.tex}

\subfile{chapters/chapter-conclusion.tex}


\backmatter

\begingroup
\printbibliography
\endgroup

\end{document}

%% file: preamble.tex
\usepackage{lmodern}        
\usepackage[T1]{fontenc}    
\usepackage[utf8]{inputenc} 

\usepackage{amsmath}    
\usepackage{amssymb}    
\usepackage{mathtools}  
\usepackage{microtype}  
\usepackage[inline]{enumitem} 
\usepackage{multirow}   
\usepackage{booktabs}   
\usepackage{subcaption} 
\usepackage[ruled,linesnumbered,algochapter,titlenumbered]{algorithm2e} 
\usepackage[usenames,dvipsnames,table]{xcolor} 
\usepackage{nag}       
\usepackage{todonotes} 

\usepackage{amsthm}
\usepackage{subfiles}
\usepackage{framed}
\usepackage{xspace}
\usepackage{multicol}
\usepackage{cancel}
\usepackage[export]{adjustbox}
\usepackage{placeins}
\usepackage{csquotes}

\usepackage{siunitx}
\usepackage[backend=bibtex,style=alphabetic,sorting=nty,maxbibnames=99,isbn=false]{biblatex}
\usepackage{xurl}

\bibliography{references}

\DeclareFieldFormat[article,inbook,incollection,inproceedings,patent,thesis,unpublished]{title}{#1}

\renewbibmacro*{doi+eprint+url}{%
    \printfield{doi}%
    \newunit\newblock%
    \iftoggle{bbx:eprint}{%
        \usebibmacro{eprint}%
    }{}%
    \newunit\newblock%
    \iffieldundef{doi}{%
        \usebibmacro{url+urldate}}%
        {}%
    }

\usepackage{hyperref}[hypertexnames=false]  
\usepackage[acronym,toc]{glossaries} 

\newcommand{\authorname}{Michael Kiran Huber} 
\newcommand{\thesistitle}{How quickly can you pack short paths?} 

\hypersetup{
    pdfpagelayout   = TwoPageRight,           
    linkbordercolor = {Melon},                
    pdfauthor       = {\authorname},          
    pdftitle        = {\thesistitle},         
    pdfsubject={Graph Theory, Disjoint Paths, Bounded Length Paths},
    pdfkeywords={Short Path Packing, Search-Tree Algorithm, Network Survivability, Graph-Sparsity Measures, Fixed-Parameter Tractability, Algorithmic Optimization, Heuristic Improvements, Computational Complexity},
    linktoc			= all,     				
}

\setpnumwidth{2.5em}        
\setsecnumdepth{subsection} 

\nonzeroparskip             
\setauthor{}{\authorname}{}{male}
\setadvisor{Univ.Prof.\ Dipl.-Inform.\ Dr.rer.nat.}{Martin Nöllenburg}{}{male}

\setfirstassistant{Univ.Lektor Dipl.-Inform.\ Dr.rer.nat.}{Manuel Sorge}{}{male}
\setsecondassistant{Univ.Ass.\ Dipl.-Ing.}{Alexander Dobler}{}{male}

\setfirstreviewer{Pretitle}{Forename Surname}{Posttitle}{male}
\setsecondreviewer{Pretitle}{Forename Surname}{Posttitle}{male}

\setsecondadvisor{Pretitle}{Forename Surname}{Posttitle}{male} 

\setregnumber{XXXXXXXX}
\setdate{01}{12}{2022} 
\settitle{\thesistitle}{\thesistitle} 
\setsubtitle{Engineering a search-tree algorithm for disjoint s-t paths of bounded length}{Engineering a search-tree algorithm for disjoint s-t paths of bounded length} 

\setthesis{bachelor}
\setcurriculum{Software and Information Engineering}{Software und Information Engineering} 

\setfirstreviewerdata{Affiliation, Country}
\setsecondreviewerdata{Affiliation, Country}

\newcommand{\onlyinsubfile}[1]{#1}
\newcommand{\notinsubfile}[1]{}

\newcommand{\figpath}{./figures/}

\newcommand{\SPP}{\textsc{Short Path Packing}\ }
\newcommand{\SPp}{\textsc{Short Path Packing}}


%% file: chapters/title.tex
\renewcommand{\onlyinsubfile}[1]{}
\renewcommand{\notinsubfile}[1]{#1}

\frontmatter 

\addtitlepage{english} 

\begin{abstract}
We study the \SPP problem which asks, given a graph $G$, integers $k$~and~$\ell$, and vertices $s$ and $t$,  whether there exist $k$ pairwise internally vertex\mbox{-}disjoint $s$-$t$ paths of length at most $\ell$. The problem has been proven to be NP-hard and fixed\mbox{-}parameter tractable parameterized by $k$ and $\ell$. 
While most previous research on this problem has been theoretical in nature, there do exist practical approaches, such as a polynomial-time heuristic. 
However, to the best of our knowledge, no
implementation of an exact algorithm for this problem including an experimental evaluation was ever published. Therefore, in this thesis we present a new FPT-algorithm based on a search-tree approach in combination with greedy localization. 
While its worst case runtime complexity of $(k\cdot \ell^2)^{k\cdot \ell}\cdot n^{O(1)}$ is larger than the state of the art, the nature of search-tree algorithms allows for a broad range of potential optimizations.
We exploit this potential by presenting techniques for appropriate preprocessing of input graphs, for detecting trivial instances, for recognizing infeasible instances in the search tree early on and for choosing promising subproblems for finding a solution. 
Those approaches were then implemented and heavily tested on a large dataset of diverse graphs. The results show that our heuristic improvements are very effective and that for the majority of instances, we can achieve fast runtimes.
  
\end{abstract}

\begin{acknowledgements*}
First of all, I would like to thank my co-advisors Manuel Sorge and Alexander Dobler for their guidance throughout the whole process of writing this thesis. Your expertise, patience and helpful feedback were invaluable to me, and I always felt like you were genuinely curious and interested in the thesis yourself, which fueled my motivation even further. 
I also want to express my profound gratitude to my advisor and BHons mentor Martin Nöllenburg. 
His passion for his field and his engaging teaching style sparked my interest in this fascinating area of study and having been able to experience research first-hand already in my bachelor studies is a privilege that I owe to him.

I am also grateful to my friends, especially Christoph and Laurenz for making the bachelor studies such an enjoyable experience and Thomas for being a great companion during our BHons year. 
Most importantly, I want to thank my girlfriend Valentina. 
Thank you for all your love and support, and for understanding when I got lost in work. Ti amo. 

Finally, I want to dedicate this thesis to my parents, Christian and Satya. 
Thank you for creating the most supportive environment in which I could always pursue my interests and passions, I do not take it for granted.
\end{acknowledgements*}

\selectlanguage{english}

\tableofcontents 

\mainmatter

%% file: chapters/chapter-introduction.tex
\onlyinsubfile{\setcounter{chapter}{1} \newcommand{\figpath}{../figures/}}
\chapter{Introduction}

As our whole world relies on connectivity and therefore on the availability of the underlying networks, fail safety and redundancy are more important than ever. One way to quantify to what extent a network meets those conditions is the concept of \textit{network survivability}, which is defined as ``the ability of a network to maintain its communication in the face of equipment failure'' \cite{SoniEtAl1999}. When we model a network by a graph, the survivability can for example be measured by considering the number of disjoint connections between communication nodes in the network. A high number of such disjoint connection paths ensures full connectivity of the network even in case of multiple failures.

In many applications, however, the mere existence of alternative disjoint paths still might not guarantee proper functionality. If redundant paths used in case of failure involve too many intermediate communication nodes, this could introduce delays in the transmission, which can negatively influence the quality of service. Therefore, an additional constraint to those disjoint connection paths can be to not exceed a certain length, ensuring transmissions with low latency.

This gives a practical motivation for the formal definition of the \SPP problem. It asks, given an undirected graph $G$, integers $k,\ell$ and vertices $s,t$, whether there exist $k$ disjoint paths that each have a length of at most $\ell$ and go from $s$ to $t$. In general, disjoint here may mean edge-disjoint or vertex-disjoint – in this thesis, however, we are concentrating on vertex-disjointness, as both cases are easily reducible to each other. This problem does not have a consistent name throughout the literature and can also be found as \textsc{Single-Commodity Hop Constrained Survivable Network Design} \cite{Fritz2011}, \textsc{Bounded Vertex-Disjoint Paths} \cite{GolovachThilikos2011}, or \textsc{Node-Disjoint Length-Restricted Paths} \cite{Bley1997}. In this thesis, for succinctness we use \SPp.

Another motivation, coming from a more theoretical perspective, is the area of graph-sparsity measures, which describe structural properties of graphs and play a large role in developing efficient algorithms. One sparsity measure of interest is the concept of so-called \textit{$r$-admissibility} \cite{Dvorak2013}, which involves determining the number of vertex-disjoint paths of length at most some integer $r$ between two vertices. Therefore, an algorithm for \SPP can also be used to calculate the $r$-admissibility of a graph.

\section{Related Work}

The problem of finding disjoint $s$-$t$ paths in a graph is a classical problem in the field of discrete algorithms. 
One of the earliest results concerning this topic is Menger's theorem \cite{Menger1927}, which states an equality between the maximum number of edge-/vertex-disjoint paths and the size of a minimum edge/vertex cut. 
A generalization of finding the maximum number of disjoint paths is finding a maximum flow in a flow network. The max-flow min-cut theorem by Ford and Fulkerson establishes an equality between maximum flows and minimum edge cuts \cite{FordFulkerson1956}. 
Adámek and Koubek showed that the max-flow min-cut theorem does not hold for flows in which the length of flow paths is bounded \cite{AdamekKoubek1971}. 
Later, Lovász et al.\ showed that Menger's theorem does not directly generalize to length-bounded paths. For the cases $\ell=2,3,4$, where $\ell$ is the maximum length allowed, a similar relation for a modified cut definition holds, while for $\ell \geq 5$ they gave upper and lower bounds on the relation \cite{LovaszEtAl1978}.

Regarding the complexity of the problem, Itai et al.\ provided efficient algorithms for the vertex-disjoint cases where $\ell \leq 4$ and proved NP-hardness of the problem for $\ell \geq 5$ \cite{ItaiEtAl1982}. Baier et al.\ later showed that the problem is even hard to approximate \cite{BaierEtAl2010}. On the other hand, when considering the parametrized complexity, according to Golovach and Thilikos the problem is fixed-parameter tractable parametrized by $k$ and $\ell$ \cite{GolovachThilikos2011}, which means that there exists an algorithm which has a runtime of $f(k,\ell) \cdot n^{O(1)}$. For small $k$ and $\ell$, this is significantly better than the trivial XP algorithm with runtime $O(n^{k\cdot\ell})$ which just tries out all possible $k\cdot\ell$-size vertex sets.

A heuristic for the corresponding optimization problem maximizing $k$, including an experimental evaluation, was given by Perl and Ronen, using a solution augmentation approach \cite{RonenPerl1984}. An ILP formulation is provided by Bley in his diploma thesis \cite{Bley1997}.

\SPP should not be confused with the very similar problem of determining $k$ disjoint $s$-$t$ paths with a total length of at most $L$, which has been shown to be solvable in polynomial time \cite{Suurballe1974, SuurballeTarjan1984}.

In the area of developing exact algorithms, to our knowledge, we are only aware of two works. In his diploma thesis, Bley described an enumeration algorithm for the problem which involves branching over all possible $\ell$-bounded $s$-$t$ paths \cite{Bley1997}. The first FPT-algorithm we know of was given by Golovach and Thilikos, who used a color-coding technique, achieving a runtime of $2^{O(k\cdot\ell)}\cdot m \cdot \log n$.

There exist many other publications around this problem and generalizations of it, e.g.\ involving multiple terminal pairs or even a general terminal set \cite{BelmonteEtAl2021}, restricting the paths to only shortest paths \cite{Eilam-Tzoreff1998}, considering edge-weights instead of unit length edges \cite{LiEtAl1992} and work focussing on practical networking applications \cite{SrinivasModiano2003, XiaoEtAl2006}.

\section{Our Contribution}

As mentioned above, the only FPT algorithm we know of uses a color-coding technique. Because the application of color-coding involves dynamic programming, a straight-forward implementation requires building the necessary tables of exponential size with regards to to the parameters, which in fact means that the average runtime is always very close to the worst case runtime – therefore, there is also not much optimization potential regarding heuristic improvements. 

An algorithmic paradigm that is better suited for tailormade heuristic improvements and optimizations are search-tree algorithms, which are based on recursively branching over a set of partial solutions, generating restricted child instances of the problem. There, on one hand we can usually decrease the runtime by cleverly recognizing child instances which can never yield a solution and on the other hand by finding good strategies for choosing promising subproblems to expand towards a solution. A common technique applied to achieve FPT runtime in search-tree algorithms is the concept of greedy localization \cite{DehneEtAl2004}. There, the branching exploits some properties of greedily computed solutions, which have bounded size and thus we can achieve a bounded branching factor.

Therefore, for a start we present an alternative proof that the problem is in FPT by providing an algorithm with a runtime of $(k\cdot\ell^2)^{k\cdot\ell} \cdot O(n + m)$. While this runtime is larger than the one shown be Golovach and Thilikos in the worst case, it builds on a search-tree technique which, as mentioned, possibly allows for faster runtimes by optimizations.

We then extend and speed up the search-tree approach in multiple ways: First, we give techniques for preprocessing the graph to reduce its size and existing algorithms for detecting trivial instances before even starting the traversal of the search tree. For the search tree itself, we present techniques for early detection of the infeasibility of subproblems and detection of symmetries, which allow for pruning the search tree. Furthermore, we also discuss potential strategies for choosing promising subproblems to expand next. 

Finally, we implement the algorithm and explain the details of doing so, before we finally conduct an experimental evaluation based on this implementation. There, we aim on one hand to show the effectiveness of the several improvements used and on the other hand to show the performance of the complete algorithm for different graphs and parameters. In the end, we discuss and analyze the results of the experiments, which show to be very promising.

\section{Structure of the Thesis}

This thesis is divided into seven chapters. The following Chapter 2 contains preliminary definitions together with the notation used throughout the thesis. In Chapter 3, we formally define the problem, give the approach for our search-tree algorithm and prove its correctness and runtime. Furthermore, we introduce a broad range of heuristic improvements for the search-tree approach in Chapter 4. To conduct an experimental evaluation, we first give insights into the details of our implementation of the algorithm in Chapter 5, before finally presenting the results of its evaluation in Chapter 6, followed by concluding remarks in Chapter~7.

%% file: chapters/chapter-preliminaries.tex
\onlyinsubfile{\setcounter{chapter}{1}}
\chapter{Preliminaries}
\label{ch:prelim}

By $\mathbb{N}$, we denote the set of all natural numbers starting at 1 and for any $i \in \mathbb{N}$,  $[i] = \{1,\ldots,i\}$. For $i,j \in \mathbb{N}$, $[i,j] = \{i,i+1,\ldots,j-1,j\}$. We write $(a_i)_{i \in [k]}$ to denote a list of elements $(a_1, a_2, \ldots, a_k)$. For such a list $A = (a_i)_{i \in [k]}$, $A[i]$ denotes the $i$th element $a_i$ of the list. Whenever we write that we \textit{insert element $x$ in list $A$ at index $i'$}, the result is a new list $A'$ with $(a_1, a_2, \ldots, a_{i'-1}, x, a_{i'}, \ldots, a_k)$.

\section{Graph Theory}
\vspace{-10pt}
\begin{figure}[hbt]
\centering
  \includegraphics[width=.9\textwidth]{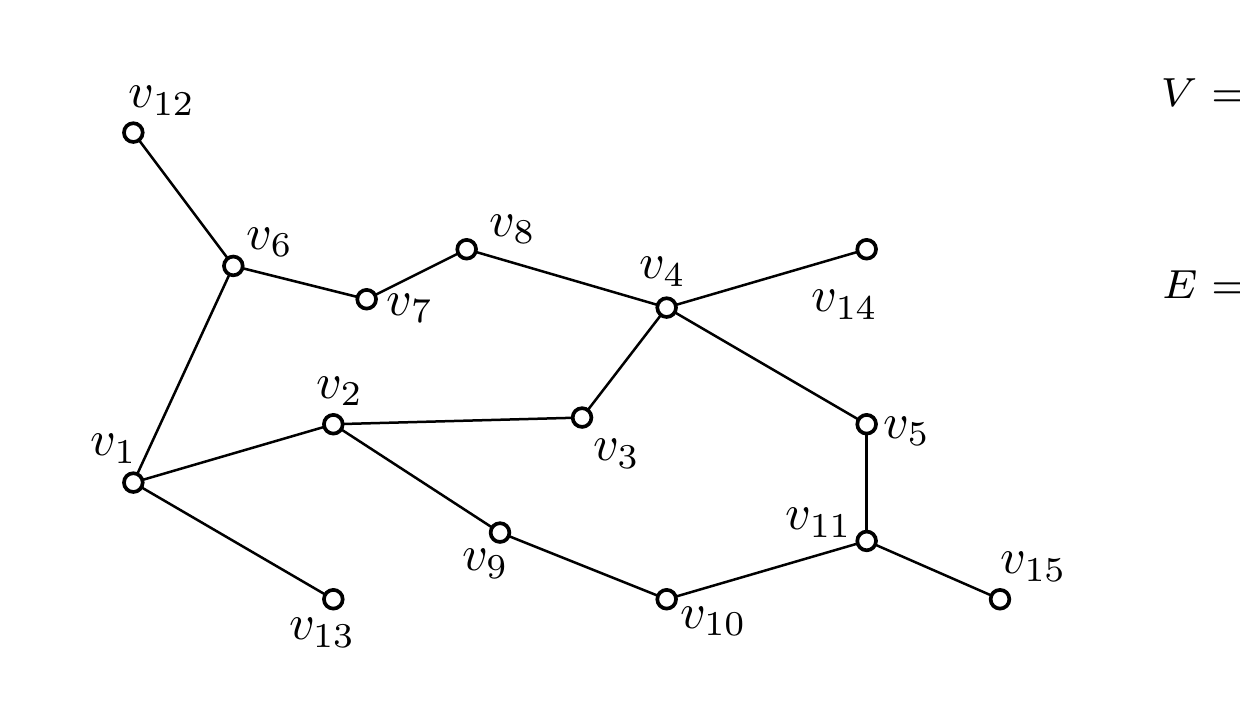}
  \vspace{-10pt}
    \caption{Example of a graph and its corresponding vertex set $V$ and edge set $E$.}
    \label{fig:prel:exgr}
\end{figure}

A graph $G = (V,E)$ is a tuple consisting of the set of \textit{vertices} $V$, also denoted as $V(G)$ and the set of \textit{edges} $E \subseteq \{\{u,v\} \mid u,v \in V \text{ and } u\neq v\}$, also denoted as $E(G)$. As an edge is given by a set of two vertices, the order does not matter. For this reason, we call those types of graphs \textit{undirected} graphs. For \textit{directed} graphs, the order of an edge matters, i.e. $E \subseteq V \times V$. If not specified otherwise, we only consider undirected graphs. Two vertices $u,v$ are \textit{adjacent}, iff there exists an edge $\{u,v\} \in E$. An edge $e$ is incident to a vertex $u$, iff $u \in e$. 

The \textit{induced subgraph} $G[V']$ on the vertex set $V' \subseteq V$ is the graph given by considering only the vertices in $V'$ and their mutual edges – formally, $G[V'] = (V', \{\{u,v\} \mid u,v \in V', \{u,v\} \in E\})$. By $G \setminus V'$ we denote the graph obtained be removing all vertices in $V'$ and their incident edges, i.e.\ the induced subgraph $G[V \setminus V']$. 

A path $P$ is a non-empty sequence of distinct, consecutively adjacent vertices. As a path can also be considered a graph, we denote by $V(P)$ the vertices and by $E(P)$ the edges of some path $P$. The length $len(P)$ of a path is given by $|V(P)|-1$, i.e. the number of vertices in the path minus one. We call a path an $s$-$t$ path if the first vertex of the path is $s$ and the last is $t$. The \textit{internal vertices} of an $s$-$t$ path $P$ are given by $V(P) \setminus \{s,t\}$, namely every vertex in the path but $s$ and $t$. The \textit{distance} of two vertices $u$ and $v$, $dist(u,v)$ is given by the length of the shortest $u$-$v$ path. In particular, $dist(v,v) = 0$.

We call a collection of $n$ paths $(P_i)_{i \in [n]}$ \textit{vertex-disjoint}, iff for any $i,j \in [k]$ with $i\neq j$ it holds that $V(P_i) \cap V(P_j) = \emptyset$, i.e. no two paths contain the same vertex. The paths are called \textit{edge-disjoint}, iff for any $i,j \in [k]$ with $i\neq j$ it holds that $E(P_i) \cap E(P_j) = \emptyset$, i.e. no two paths use the same edge. A collection of $k$ $s$-$t$ paths $(P_i)_{i \in [k]}$ is called \textit{internally vertex-disjoint}, iff for any $i,j \in [k]$ with $i\neq j$ it holds that $V(P_i) \cap V(P_j) = \{s,t\}$, i.e. the paths share no vertices except their endpoints. In this thesis, if not mentioned otherwise, whenever we write \textit{disjoint}, we mean \textit{internally vertex-disjoint}.

If the graph $G$ has the property that every pair of vertices $(u,v)$ is connected by a path, then we call the graph \textit{connected}, otherwise we call it \textit{disconnected}. A disconnected graph consists of multiple \textit{connected components}, which are the distinct subgraphs of the graph in which every pair of vertices is connected. A \textit{minimum $s$-$t$ vertex separator} (sometimes called \textit{vertex cut}) is the smallest set of vertices that, if removed, disconnect the graph, such that no $s$-$t$ path exists, i.e. $s$ and $t$ belong to different connected components of the disconnected graph. A \textit{minimum $s$-$t$ edge cut} similarly is the smallest set of edges such that upon removal, $s$ and $t$ are disconnected.

By $N_r(v)$ we denote the \textit{$r$-neighborhood} of a vertex $v$. This is given by the set of vertices $u$ which have a distance of at most $r$ from $v$, i.e. $N_r(v) = \{u \mid dist(v,u)\leq r\}$. As $dist(v,v)=0$, this implies that $v \in N_r(v)$ for any $r$.

\section{Algorithmic Techniques}

\subsection{Greedy Algorithms}
An algorithm is called a \textit{greedy algorithm} if it can be characterized by always making locally optimal choices, i.e.\ when presented with multiple choices, it always selects the best available choice without considering possible future consequences \cite{Black2005, EncyclopediaOfMathematics}. While in some cases greedy algorithms do give optimal solutions as a result, in many cases they do not.
For optimization problems, where we want to maximize or minimize some target function, greedy algorithms in general give valid solutions to the problem which might not be optimal, but still good enough for practical use. For decision problems on the other hand, where we want to know whether \textit{any} solution satisfying the given constraints exists, the situation looks a bit different. Here, if lucky, a greedy algorithm outputs a solution satisfying the constraints allowing us to answer the decision problem, but if the greedy solution does not fulfill the constraints, no statement about whether the given instance is a yes- or a no-instance can be made.

\subsection{Search Trees and Backtracking}

\begin{figure}[t]
  \centering
  \includegraphics[page=2]{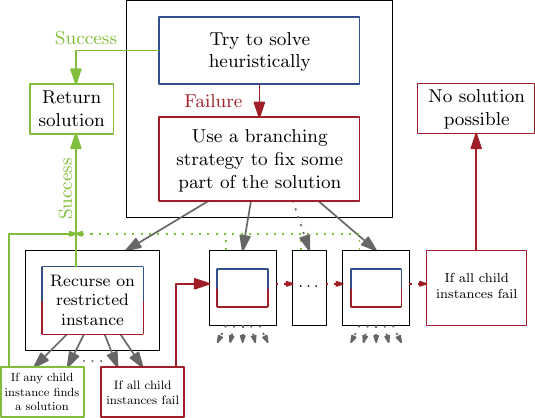}
  \caption{Illustration of the general procedure when employing a search-tree technique for solving a problem.}
\end{figure}

For NP-hard problems, for which we assume that in general no efficient, polynomial-time algorithms exist, one simple algorithmic paradigm for finding a solution is trying out all possible solutions. If this is done in an enumerative, exhaustive way going through every possible solution like trying to crack a bicycle lock by trying all possible combinations, this is called \textit{brute force} search. The set of all possible solutions is often called the \textit{solution space}, and even for a brute force approach one has to think of a way to traverse the solution space exhaustively - and often there are multiple options to do so. A refinement of the brute force approach is given by \textit{search-tree algorithms}, also called \textit{backtracking algorithms} \cite{Gurari1999}, which perform the search of the solution space in a more systematical way.

The name \textit{search-tree algorithm} already gives the hint that the search will be performed in some tree-structured way. In fact, the root of the search tree is given by an empty solution candidate. Then, we generate child candidates where we fix one part of the solution recursively until we either find a solution or, based on some criteria, can be sure that expanding this branch further can never yield a solution. In that case, we \textit{backtrack} back to the parent and try to find our solution in another branch.

\subsection{Greedy Localization}

Highly relevant for this thesis is the concept of \textit{Greedy Localization} \cite{DehneEtAl2004}. In its core, it is a technique (mostly used for maximization problems) employing greedy strategies within a search-tree approach to limit the number of choices to branch over.

The idea is, given a constrained instance in the search tree, first to try to attempt solving this instance using a greedy algorithm. With regards to a \SPP instance, where we simply want to find $k$ disjoint paths, we employ a greedy algorithm to compute disjoint paths until no more can be found. If we are able to find $k$ paths in such a way, we are successful, but if we are not, then we know that a solution needs to intersect our $k$ greedily computed paths in some way. Therefore, the partial solution set computed by the greedy attempt can give us additional information about the structure of the solution, which allows us to ``localize'' our search by narrowing down the possibilities.

%% file: chapters/chapter-theoretical-results.tex
\onlyinsubfile{\setcounter{chapter}{2} \newcommand{\figpath}{../figures/}}
\chapter{Short Path Packing}
\label{ch:theor}

In this chapter, we present the theoretical results of the thesis. 
In Section \ref{ch:theor:prob-def}, we define the general problem setting and give some examples. 
Then, in Section \ref{ch:theor:algo} we look into a greedy approach and its limitations for solving the problem. Towards finding a branching strategy for a search-tree approach, we introduce a modified variant of the problem in Section \ref{ch:theor:algo:cp} before presenting our final search-tree algorithm and a proof of its correctness and its running time in Section \ref{ch:theor:exact-algo}.

\section{Problem Definition}
\label{ch:theor:prob-def}

\newcommand{\bigcupdot}{\ \cdot \hspace{-7.5pt}\bigcup}
\newcommand{\ibr}{{i_\beta}}
\newcommand{\jbr}{{j_\beta}}

The formal description of the problem at hand can be given as follows:
\begin{framed}
\textsc{Short Path Packing (SPP)}
\begin{description}
	\item \textit{Instance:} A graph $G$, two vertices $s, t \in V(G)$ and two integers $k, \ell \in \mathbb{N}$.
	\item \textit{Question:} Are there $k$ internally vertex-disjoint $s$-$t$ paths  in $G$, each of length at most $\ell$?
\end{description}
\end{framed}

Clearly, for $k=1$, the problem just reduces to finding the shortest $s$-$t$ path in $G$ and checking whether its length exceeds $\ell$, which is feasible in polynomial time. For larger~$k$, Itai et al.\ showed that for values greater than 2, no such polynomial time algorithm is likely to exist:

\begin{theorem}[\cite{ItaiEtAl1982}]
	\SPP is NP-complete for $k \geq 2$ and $\ell \geq 5$.
\end{theorem}

In their work, at the beginning they also disregard the cases $\ell = 1,2$ as trivial.
As we are considering only simple graphs without multi-edges, any instance with $\ell=1$ can only be satisfied if $k=1$, reducing it to checking whether $s$ and $t$ are adjacent. For $\ell=2$, the problem just boils down to counting common neighbors of $s$ and $t$. For the cases $\ell=3,4$, Itai et al.\ present polynomial-time algorithms that are based on reducing the problem to a maximum matching instance. For $\ell \geq 5$, however, they are able to show NP-completeness by giving a polynomial-time reduction from the \textsc{3-SAT} problem.

As we are mainly concerned about engineering a new algorithm for solving hard instances, from now on we therefore only concentrate on instances where $k \geq 2$ and $\ell \geq 5$.

In our case, \SPP is formulated as a \textit{decision problem}, meaning that the answer to an instance is just ``yes'' or ``no''. But of course in the case of a yes-instance, there has to exist a \textit{solution}, which is a collection of internally vertex-disjoint $s$-$t$-paths $\mathcal P = (P_i)_{i \in [k]}$, such that for any $P_i$, it holds that $len(P_i) \leq \ell$. Many publications also focus on the \textit{optimization problem}, that is given some length bound and two vertices, maximize the number $k$ of disjoint length-bounded paths. In this thesis, however, we are focussing on the decision problem, as an algorithm for the decision problem can always also be employed for solving the optimization problem.

\begin{figure}[t]
  \centering
  \includegraphics[width=0.6\textwidth]{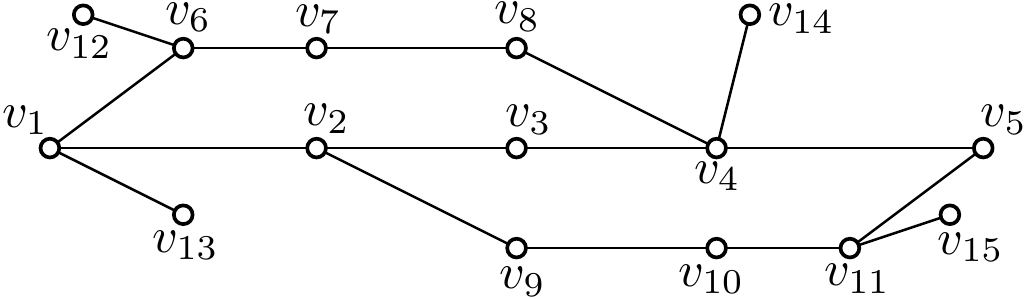} \\
  $I = (G, k = 2, \ell=5, s = v_1, t = v_5)$
  \caption{Example instance $I$ for \SPP based on the same graph as displayed in Figure \ref{fig:prel:exgr}.}
  \label{fig:theor:example}
\end{figure}

\begin{figure}[t]
  \centering
  \includegraphics[width=0.6\textwidth, page=2]{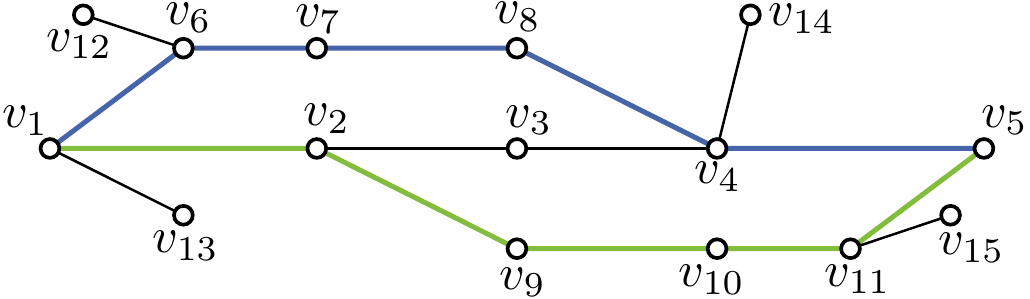} \\
 $\mathcal P = ((v_1, v_6, v_7, v_8, v_4, v_5), (v_1, v_2, v_9, v_{10}, v_{11}, v_{5}))$ 
  \caption{A set of internally vertex-disjoint $v_1$-$v_5$ paths forming a solution for the instance $I$.}
  \label{fig:theor:solution-ex}
\end{figure}

Figure \ref{fig:theor:example} shows an example instance based on the graph from Figure \ref{fig:prel:exgr} shown in the preliminaries. In this instance, if one would greedily use e.g.\ the shortest path in the middle via vertices $v_2$ and $v_3$, no further vertex-disjoint path would exist in the graph. Figure \ref{fig:theor:solution-ex} shows the solution to this instance.

\section{Greedy Approach}
\label{ch:theor:algo}
\label{ch:theor:algo:greedy}

Often, a viable starting point  is to look at a greedy algorithm for solving the problem and at the structure of the solution in the case it fails.
For \SPP and an instance $(G, k, \ell, s, t)$, a possible greedy algorithm would compute the shortest $s$-$t$~path, remove it from the graph and then repeat in the modified graph. 
As soon as the shortest path retrieved does not exist or is longer than $\ell$, we can stop. 
We obtain a solution, if we are able to retrieve $k$ paths in such a way. 
The approach is given as pseudocode in Algorithm \ref{alg:greedySPP}.

\begin{algorithm}
	\KwIn{A graph $G = (V,E)$, vertices $s$ and $t$, integers $k$ and $\ell$}
	\KwOut{$P = (P_1,P_2,...,P_k)$ – a list of $k$ disjoint paths in $G$ with a length of at most $\ell$; or $\bot$}
	\ForEach{$i \in [k]$}{
			$G_i \gets G \setminus \left(\bigcup_{j\in [i-1]} V(P_j) \setminus \{s,t\} \right)$\;
			$P_i \gets $ shortest path from $s$ to $t$ in $G_i$ or $\bot$ if none exists\;
			\If{($P_i = \bot) \lor (len(P_i) > \ell)$ }{
				\Return $\bot$\;
		}
	}
	\Return $(P_j)_{j \in [k]}$\;
	\caption{GreedySPP}	
	\label{alg:greedySPP}
\end{algorithm}

We remark, that this is not a ``classical'' greedy algorithm in the way that it finds paths until none exist. In fact, such a strategy would be viable for the corresponding optimization problem – but as we are concentrating on the decision problem, our algorithm only needs to find $k$ paths, even if there would exist more. Still, to this regard the applicability of the greedy algorithm is limited. If it returns a solution, it is a valid one, as it was obtained by repeatedly finding $s$-$t$ paths of length at most $\ell$ and then removing their internal vertices from the graph until $k$ such paths were found – the disjointness therefore is guaranteed by the fact that as soon as a vertex is used by a path, it gets removed from the graph. When the algorithm fails, however, no statement about the instance can be made. Because of this, the algorithm is sound (if it returns a result, we can rely on it) but not complete (there are cases in which the algorithm does not return any result). Our knowledge in such cases is restricted to the fact that the greedy algorithm failed, however this does not tell us whether a solution can exist or not. 

The running time of the algorithm, under our assumption that the edges have no weights, is linear with respect to the number of vertices and edges using breadth-first-search to compute shortest paths. We perform at most $k$ iterations, leaving us with a runtime of $k\cdot O(n+m)$.

Now, let $\ibr$ denote the value of the loop index $i$ in Algorithm \ref{alg:greedySPP} in the iteration at which the algorithm fails and returns $\bot$. For this situation, we can state the following observation:

\begin{observation}
	\label{obs:greedySPPreturnIteration}
	If Algorithm \ref{alg:greedySPP} returns $\bot$ in iteration $\ibr$, this means that in the graph $G_\ibr$ there exists no $s$-$t$ path of length at most $\ell$.
\end{observation}

We will come back to this observation shortly. As explained before, if the greedy algorithm returns at least $k$ paths, we are done. 
Therefore, the question is how to deal with the case of failure.  
What can we derive from the fact that our greedy algorithm did not find a solution with $k$ paths? 
Considering a no-instance of \SPp, we know that any valid algorithm which is sound and complete will fail to find a solution, thus we can accept the failure of our greedy algorithm in this case.
Therefore, the main point of interest for us are yes-instances where the greedy algorithm fails.
Intuitively, the failure of the greedy algorithm tells us that some of the previously computed paths must have been wrong. If the path $P_\ibr$, which we failed to find, does exist, it has to use at least one vertex that was already consumed by greedily determining one of the previous paths. This motivates the following lemma:

\begin{lemma}
	\label{lemma:optimalUsesGreedyVertex}
	Given an SPP instance $I = (G, k, \ell, s, t)$, assume Algorithm \ref{alg:greedySPP} failed in iteration $\ibr$ of the outer loop. Let $\mathcal P$ be a collection of $\ibr-1 < k$ disjoint $s$-$t$ paths $(P_i)_{i \in [\ibr-1]}$ of length at most $\ell$ greedily computed up to the point of failure. 
	If $I$ is a yes-instance and thus a solution $\mathcal P^* = (P^*_i)_{i \in [k]}$ exists, then the path $P^*_{\ibr}$ must use some internal vertex of the greedily computed collection of paths $\mathcal P$. 
	In formulas,
	\[ V(P^*_{\ibr}) \cap \bigg( \bigcup_{i \in [\ibr - 1]} V(P_i) \setminus \{s,t\} \bigg)  \neq \emptyset.\]
	
	\begin{proof}
		Assume, the intersection was empty. 
		This means that the path $P^*_{\ibr}$ is contained in the graph $G_{\ibr}$ (see line 2 of Algorithm \ref{alg:greedySPP}). 
		By definition, $P^*_{\ibr}$ is an $s$-$t$ path of length at most $\ell$. 
		However, by Observation \ref{obs:greedySPPreturnIteration} no $s$-$t$ path of length at most $\ell$ can exist in $G_{\ibr}$. 
		This is a contradiction, thus the intersection cannot be empty.
	\end{proof}
\end{lemma}

\section{Checkpoints}
\label{ch:theor:algo:cp}
Lemma \ref{lemma:optimalUsesGreedyVertex} naturally suggests the use of a branching mechanism to exploit the property given by the lemma: When the greedy algorithm fails to find path $P_{\ibr}$, if a solution $(P^*_i)_{i \in [k]}$ exists, the path $P^*_{\ibr}$ has to use some internal vertex of the greedily determined paths $(P_i)_{i \in [\ibr -1 ]}$. Therefore, to find a solution we could branch over all those vertices $v$ generating new instances which constrain path $P_{\ibr}$ to having to use vertex $v$ and then again try our greedy approach recursively.
But before we can do so, we need a formal notion of ``\textit{path $P$ has to use vertex $v$}''. 
One possible approach would be to use a multi-terminal formulation and split the $s$-$t$ path $P_{\ibr}$ into two halves, an $s$-$v$ path and an $v$-$t$ path. However, this comes with a large guessing overhead, as we do not know how the length constraint $\ell$ distributes over the two new subpaths. Therefore, we would need to try all possible combinations of lengths. This way, instead of creating \textit{one} child instance which encodes the constraint that path $P_\ibr$ has to use vertex $v$, we would need to create $\ell$ child instances: The first for an $s$-$v$ path of length at most 1 and an $v$-$t$ path of length at most $\ell-1$, the next one for $2$ and $\ell-2$, and so on.

To avoid this, we use a different formulation which keeps us from having to guess more than needed. We do so by extending our notion of terminal pairs (in our case $(s,t)$) to so-called terminal lists. For example, the formalization of an $s$-$t$ path using the vertex $v$ would be an $(s,v,t)$-path. We call the elements, that are neither first nor last of the list, \textit{checkpoints}. Generally speaking, for any terminal list $L_i$, an \textit{$L_i$-path} starts in the first element of the list, visits every checkpoint in $L_i$ in the corresponding order and ends in the last element of the list.

This allows us to define a modified version of \SPp: 
\begin{framed}
\textsc{Short Path Packing with Checkpoints (SPPC)}
	\begin{description}
		\item \textit{Instance:} A graph $G$, two terminal vertices $s, t \in V(G)$, two integers $k, \ell \in \mathbb{N}$, $k$~terminal lists $L_1,\ldots,L_k$ such that for each $L_i$, the first element of the list is $s$, the last element is $t$.
		\item \textit{Question:} Are there $k$ internally-vertex disjoint $s$-$t$ paths $P_1,\ldots,P_k$ in $G$, each of length at most $\ell$, such that each path $P_i$ is an $L_i$-path?
	\end{description}
\end{framed}

 Again, in the case of a yes-instance, there exists a \textit{solution}, namely a collection of paths $(P_i)_{i \in [k]}$, such that each $P_i$ is an $L_i$-path with $len(P_i) \leq \ell$. Because each $L_i$ starts with $s$ and ends with $t$, each path also is an $s$-$t$ path. Furthermore, as each $P_i$ is an $L_i$-path, it consists of subpaths $(Q_{i,j})_{j \in [|L_i|-1]}$. By $Q_{i,j}$ therein we denote the subpath going from $L_i[j]$ to $L_i[j+1]$. In such a solution $(P_i)_{i \in [k]}$, we call $\bigcup_{i \in [k]} V(P_i) \setminus L_i$ the \textit{non-terminal vertices} of the solution.

Coming back to creating a branching strategy based on greedily computed paths, we need to adapt the greedy algorithm to incorporate the new notion of terminal lists. As we now have a fixed number of (possibly distinct) terminal lists, the greedy algorithm cannot be formulated as general as Algorithm \ref{alg:greedySPP}. 
In particular, as we now not only compute one full $(s,t)$-path after one another, but subpaths between consecutive checkpoints, this changes the failure conditions – we now additionally need to keep track of the total length of the $s$-$t$ path that is currently being searched.

\begin{algorithm}
	\KwIn{A graph $G = (V,E)$, vertices $s$ and $t$, integers $\ell$ and $k$ and vertex terminal lists $(L_i)_{i\in [k]}$}
	\KwOut{$P = (P_1,P_2,...,P_{k'})$ – a collection of $k$ disjoint paths in $G$ with a length of at most $\ell$ such that\ each $P_i$ is an $L_i$-path; or $\bot$.}
	\ForEach{$i \in [k]$}{
			$\ell_i \gets 0$\;
			$G_i \gets G \setminus \big(\bigcup_{x\in [i-1]} V(P_x) \setminus \{s,t\} \big)$\;
		\ForEach{$j \in [|L_i|-1]$}{
			$G_{i,j} \gets \left(G_i \setminus \big(\bigcup_{y \in [j-1]} V(Q_{i,y}) \cup \bigcup_{x \in [k]} L_x \big)\right) \cup \{L_i[j], L_i[j+1]\} $\;
			$Q_{i,j} \gets$ shortest path from $L_i[j]$ to $L_i[j+1]$ in $G_{i,j}$ or $\bot$ if none exists\;
			\If{($Q_{i,j} = \bot) \lor (\ell_i + len(Q_{i,j}) > \ell)$ \label{line:failConditionSPPC}}{
				\Return $\bot$;
			}
			$\ell_i \gets \ell_i + len(Q_{i,j})$\;
		}
		$P_i \gets (Q_{i,j})_{j \in [|L_i|-1]}$\; 
	}
	\Return $(P_j)_{j \in [k]}$\;

	\caption{First approach for GreedySPPC}	
	\label{alg:greedySPPC}
\end{algorithm}

Those changes are incorporated in Algorithm \ref{alg:greedySPPC}.
Towards a branching strategy we consider how we can reformulate Lemma \ref{lemma:optimalUsesGreedyVertex} to respect the changes made to the algorithm. A naive approach would be to extend Observation \ref{obs:greedySPPreturnIteration} to the claim that in the case of failure in the iteration $\ibr$ of the outer loop, no $L_\ibr$-path of length at most $\ell$ could exist in the graph $G_\ibr$. However, this claim proves to be wrong, as the search for the $L_\ibr$\nobreakdash-path~$P_\ibr$, where the failure arises, itself consists of multiple greedy steps finding subpaths connecting the checkpoints. Therefore, even though the algorithm fails in iteration $\ibr$ of the outer loop, an $L_\ibr$-path of length at most $\ell$ could still exist in $G_\ibr$ – only the algorithm might have greedily chosen a subpath connecting the first two checkpoints which consumed vertices necessary for completing the $L_\ibr$-path by the following subpaths. 

Therefore, we have to additionally focus on the iteration $\jbr$ of the inner loop where the failure happens, to correctly reason about the situation. If we know that the algorithm failed while computing the subpath $Q_{\ibr,\jbr}$ going from $L_\ibr[\jbr]$ to $L_\ibr[\jbr+1]$, it would seem to make sense to focus on only this subpath and therefore branch by inserting previously used vertices at the position between $L_\ibr[\jbr]$ and $L_\ibr[\jbr+1]$.

Furthermore, we have to reconsider our definition of ``previously used vertices'', meaning the candidate vertices that we branch over. In the first simple greedy approach of Algorithm \ref{alg:greedySPP}, we considered the failure at the ``path-level'' in the sense, that we viewed the whole path $P_{i}$ as failed. If we now switch to a more fine-grained view on the ``subpath-level'' where we consider the failure of finding a specific subpath, as briefly mentioned the failure might as well come from the fact that some predecessor subpath $Q_{i,j'}$ with $j'<j$ might occupy a vertex that is contained in the solution subpath $Q^*_{i,j}$. 

Lastly, another detail arises. In Line \ref{line:failConditionSPPC} of Algorithm \ref{alg:greedySPPC}, we can see two different conditions of failure which need to be taken care of separately: 
The first condition $Q_{i,j} = \bot$ describes the situation where some subpath $Q_{i,j}$ does not exist in the graph $G_{i,j}$, i.e. $L_i[j]$ and $L_i[j+1]$ are in different components of the graph. Let us call this \textit{Failure Condition 1 (FC1)}. In this case, we can come back to the observation made before, that the solution subpath has to use some vertex of the previously computed (sub)paths. This is formalized in Lemma \ref{lemma:optimalUsesGreedyVertexSPPC2}.

\begin{lemma}
	\label{lemma:optimalUsesGreedyVertexSPPC2}
	Given an SPPC instance $I = (G, k, \ell, s, t, (L_i)_{i\in [k]})$, assume Algorithm \ref{alg:greedySPPC} failed in iteration $\ibr$ of the outer and $\jbr$ of the inner loop because of Failure Condition~1. Let $\mathcal P$ be the returned collection of $\ibr-1$ disjoint paths $(P_i)_{i \in [\ibr-1]}$ of length at most $\ell$ such that each $P_i$ is an $L_i$-path. Furthermore, let $\mathcal Q = (Q_{\ibr,j})_{j \in [\jbr-1]}$ be the sequence of greedily computed subpaths in the iteration $\ibr$ of the outer loop. If $I$ is a yes-instance and thus a solution $(P^*_i)_{i \in [k]}$ each consisting of subpaths $(Q^*_{i,j})_{j\in[|L_i|-1]}$ exists, then the path $Q^*_{\ibr,\jbr}$ must use some non-terminal vertex of the greedily computed collection of paths $\mathcal P$ or of the set of subpaths $\mathcal Q$. In formulas,
	\[ V(Q^*_{\ibr,\jbr}) \cap \Bigg( \bigg( \bigcup_{i \in [\ibr-1]} V(P_i) \cup \bigcup_{j \in [\jbr-1]} V(Q_{\ibr,j}) \bigg) \setminus \bigcup_{i\in[k]} L_i  \Bigg) \neq \emptyset.\]
\begin{proof}
	Assume, the intersection was empty. 
	This means that the path $Q^*_{\ibr, \jbr}$ is contained in the graph $G_{\ibr, \jbr}$ (see line 3 of Algorithm \ref{alg:greedySPPC}). 
	By definition, $Q^*_{\ibr, \jbr}$ is an $L_{\ibr}[\jbr]$\nobreakdash-$L_{\ibr}[\jbr+1]$ path. 
	As we assume the algorithm to have returned due to Failure Condition 1, this implies that there exists no $L_{\ibr}[\jbr]$\nobreakdash-$L_{\ibr}[\jbr+1]$ path in the graph $G_{\ibr, \jbr}$.
	This is a contradiction, thus the intersection cannot be empty.
	\end{proof}
\end{lemma}

\begin{figure}[t]
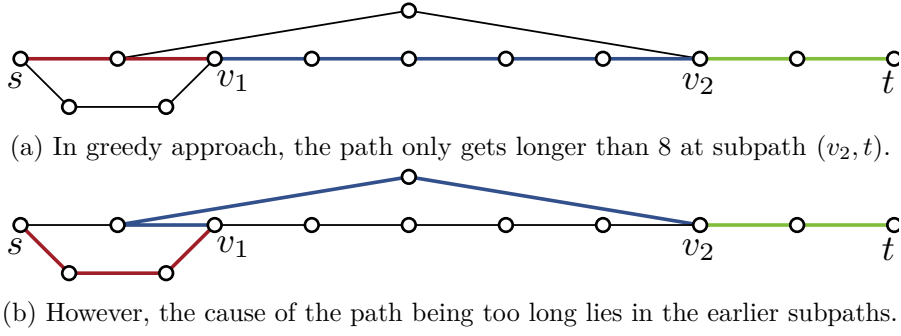

	\centering
	\begin{subfigure}[b]{0.8\textwidth}
		\centering
  		\includegraphics[width=\textwidth, page=2]{\figpath chapter-theoretical-results/FC2_example2_new.pdf}
  		\caption{In greedy approach, the path only gets longer than 8 at subpath $(v_2,t)$.}
	\end{subfigure} \hfill
	\begin{subfigure}[b]{0.8\textwidth}
		\centering
  		\includegraphics[width=\textwidth, page=3]{\figpath chapter-theoretical-results/FC2_example2_new.pdf}
  		\caption{However, the cause of the path being too long lies in the earlier subpaths.}
	\end{subfigure}
	\caption{SPPC instance where we want to find a $(s,v_1,v_2,t)$-path of length $\leq 8$.}
	\label{fig:FC2example}
\end{figure}

On the other hand, the condition $\ell_i + len(Q_{i,j}) > \ell$, which we call \textit{Failure Condition 2 (FC2)}, requires different reasoning. It describes the case when the total length of the previously computed subpaths plus the current subpath exceeds our length bound $\ell$. Here, we cannot just argue that the solution subpath $Q^*_{i,j}$ must use some previously used nonterminal vertex, as it could be the case that the path $Q_{i,j}$ is already the correct one which also is part of the solution, and the length overflow came from some incorrect predecessor subpath. Such a situation is shown in Figure \ref{fig:FC2example}. For simplification, we only depict one path – in general, the failure causes of course can be manifold. One should solely see this as a counterexample to show that the solution subpath of $(v_2,t)$ is correctly computed by the greedy algorithm, even though it fails at this point. 

The problem with our example is that by greedily taking the shortest $s$-$v_1$ path, we cut off the short $v_1$-$v_2$ path. However, at this point in the algorithm we do not know yet that this will be a problem as it not yet causes our path to exceed the maximum length. The failure only happens when the algorithm determines the $v_2$-$t$ path, which length added to the lengths of the previously found paths does in fact exceed $\ell$. Just for that reason, however, we can not ``blame'' the $v_2$-$t$ path and conclude that we need to insert a checkpoint there – our example perfectly illustrated that the problem indeed lies at the first two subpaths, namely that the path from $v_1$ to $v_2$ needs to use a vertex occupied by the path from $s$ to $v_1$.

To correctly handle this Failure Condition 2, we therefore have to slightly generalize Lemma \ref{lemma:optimalUsesGreedyVertexSPPC2}. While there we could restrict ourselves to the fact that \textit{just} the failed subpath must use some previously used non-terminal vertex, here we claim that \textit{some} subpath of the failed path must use such a previously used non-terminal vertex. This is done in Lemma \ref{lemma:optimalUsesGreedyVertexSPPC3}:

\begin{lemma}
	\label{lemma:optimalUsesGreedyVertexSPPC3}
	Given an SPPC instance $I = (G, k, \ell, s, t, (L_i)_{i\in [k]})$, assume Algorithm \ref{alg:greedySPPC} failed in iteration $\ibr$ of the outer and $\jbr$ of the inner loop because of Failure Condition~2. Let $\mathcal P$ be the returned set of $\ibr-1$ disjoint paths $(P_i)_{i \in [\ibr-1]}$ of length at most $\ell$ such that each $P_i$ is an $L_i$-path. Furthermore, let $\mathcal Q = (Q_{\ibr,j})_{j \in [\jbr-1]}$ be the sequence of greedily computed subpaths in iteration $\ibr$ of the outer loop. If $I$ is a yes-instance and thus a solution $(P^*_i)_{i \in [k]}$, each consisting of subpaths $(Q^*_{i,j})_{j\in[|L_i|-1]}$ exists, then there exists a path $Q^*_{\ibr,j'} \in (Q^*_{\ibr,j})_{j \in [\jbr]}$ which uses some non-terminal vertex of the greedily computed collection of paths $\mathcal P$ or of the collection of subpaths $\mathcal Q \setminus Q_{\ibr,j'}$. In formulas,
	\[ \bigcup_{j \in [\jbr]} \Bigg( V(Q^*_{\ibr,j}) \cap \bigg( \Big( \bigcup_{j' \in [\jbr],j' \neq j} V(Q_{\ibr,j})\cup \bigcup_{i \in [\ibr-1]} V(P_i) \Big) \setminus \bigcup_{i\in[k]} L_i \bigg) \Bigg) \neq \emptyset.\]

\begin{proof}
	Assume, the intersection was empty. 
	This means that any solution subpath $Q^*_{\ibr,j}$ for $j \in [\jbr]$ neither contains an internal vertex of another path $Q_{\ibr,j'}$ with $j' \neq j$ nor it contains a non-terminal vertex of another path $P_i$ for $i \in [\ibr-1]$. This means, that each subpath $Q^*_{\ibr,j}$ for each $j \in [\jbr]$ must be contained in the graph $G_{\ibr}$.  
	
	Now we need to consider two cases:
	\begin{description}[topsep=0pt]
		\item Case 1: $Q^*_{\ibr,j} = Q_{\ibr,j}$ for all $j \in [\jbr]$. This is a contradiction, as then the algorithm could have never failed due to FC2.
		\item Case 2: There exists at least one $j'$, such that $Q^*_{\ibr,j'} \neq Q_{\ibr,j'}$. As the total length of subpaths $\mathcal Q$ exceeds $\ell$ and the length of the solution subpaths $\mathcal Q^*$ clearly does not, there has to exist at least one $j'$ such that $len(Q^*_{\ibr,j'}) < len(Q_{\ibr,j'})$. We choose the one with minimal $j'$. By our assumption we know that $Q^*_{\ibr,j'}$ is contained in $G_{\ibr}$. However, this means that our greedy algorithm should have found $Q^*_{\ibr,j'}$ or another even shorter path instead of $Q_{\ibr,j'}$. This is a contradiction. 
	\end{description}
	As both cases lead to a contradiction, the intersection cannot be empty.
	\end{proof}
\end{lemma}

\section{Search-Tree Algorithm}
\label{ch:theor:exact-algo}

Those results now allow us to formulate the exact algorithm used for solving the problem. As briefly mentioned in previous sections, we will be using a search-tree approach.

\begin{algorithm}
	\KwIn{Instance $I = (G,k,\ell,s,t,(L_i)_{i\in [k]})$ }
	\KwOut{Empty set, if no solution exists. Otherwise, collection of $k$ disjoint $s$-$t$ paths in $G$ such that each $P_i$ is an $L_i$-path and has a length of at most $\ell$.}
	\uIf{$I$ is a trivial no-instance}{\Return \{\}\;}
	\uElse{
		$S \gets$ GreedySPPC($I$)\;
		\uIf{GreedySPPC was successful}{\Return $S$\;}
		\uElseIf{GreedySPPC failed due to FC1}{
			\Return BranchingRule1($I,S$)\;
		}
		\ElseIf{GreedySPPC failed due to FC2}{
			\Return BranchingRule2($I,S$)\;
		}
	}
	\caption{Search-Tree Algorithm for SPPC}
	\label{alg:searchTreeSPPC}
\end{algorithm}

Algorithm \ref{alg:searchTreeSPPC} summarizes the search-tree approach, about which we will go into further detail in the following section. To fully specify the algorithm, we need to clarify how to determine trivial no-instances and define the subroutine GreedySPPC and the Branching Rules 1 and 2.

Naively recognizing a trivial no-instance is straightforward: 

\begin{observation}
	\label{obs:theor:trivial}
	 An instance $I = (G,k,\ell,s,t,(L_i)_{i\in [k]})$ is a trivial no-instance, if there exist some terminal list $L_i$ which has more than $\ell+1$ elements. Then, we cannot satisfy the terminal list with any path of length at most $\ell$.
\end{observation}

Furthermore, GreedySPPC basically is a refined version of Algorithm \ref{alg:greedySPPC}. However, we now want to be able to differentiate between the two Failure Conditions and we also want to possibly return the incomplete path consisting of  all subpaths the algorithm found before it failed. This is done in Algorithm \ref{alg:greedySPPC2}.

\begin{algorithm}
	\KwIn{A graph $G = (V,E)$, integers $k$ and $\ell$ and vertex terminal lists $(L_i)_{i\in [k]}$}
	\KwOut{$\mathcal P = (P_1,P_2,...,P_{k'-1}, P_{k'}) $, where $P_{k'} = (Q_{k',1}, ..., Q_{k',j})$ – a collection of $k' \leq k$ disjoint paths in $G$ such that each $P_i$ (except $i = k'$) is an $L_i$-path of length at most $\ell$ and $P_{k'}$ is a sequence of subpaths connecting the first $j+1$ checkpoints in $L_{k'}$.}
	\ForEach{$i \in [k]$}{
			$\ell_i \gets 0$\;
			$G_i \gets G \setminus \big(\bigcup_{x\in [i-1]} V(P_x) \big)$\;
		\ForEach{$j \in [|L_i|-1]$}{
			$G_{i,j} \gets \left(G_i \setminus \big(\bigcup_{y \in [j-1]} V(Q_{i,y}) \cup \bigcup_{x \in [k]} L_x \big)\right) \cup \{L_i[j], L_i[j+1]\} $\;
			$Q_{i,j} \gets$ shortest path from $L_i[j]$ to $L_i[j+1]$ in $G_{i,j}$ or $\bot$ if none exists\;
			\uIf{$Q_{i,j} = \bot$}{
				\Return $(P_j)_{j \in [i-1]} \cup (Q_{i,j})_{j\in [j-1]}$ (FC1)\;
			} \ElseIf{$\ell_i + len(Q_{i,j}) > \ell$}{
				\Return $(P_j)_{j \in [i-1]} \cup (Q_{i,j})_{j\in [j-1]}$ (FC2)\;
			}
			$\ell_i \gets \ell_i + len(Q_{i,j})$\;
		}
		$P_i \gets (Q_{i,j})_{j \in [|L_i|-1]}$\; 
	}
	\Return $(P_j)_{j \in [k]}$\;

	\caption{GreedySPPC}	
	\label{alg:greedySPPC2}
\end{algorithm}

It is easy to see that all the Lemmas and Observations which held for Algorithm \ref{alg:greedySPPC} still hold for Algorithm \ref{alg:greedySPPC2}. The only change is the explicit distinction between the two failure conditions and the consideration of the incomplete subpaths $\mathcal Q$ into the output.

Finally, the branching rules in Algorithm \ref{alg:searchTreeSPPC} highly utilize the results from Lemma \ref{lemma:optimalUsesGreedyVertexSPPC2} and Lemma \ref{lemma:optimalUsesGreedyVertexSPPC3}. Again, let $\ibr$ denote the value of the loop index $i$ and $\jbr$ the value of the loop index $j$ in Algorithm \ref{alg:greedySPPC2} in the iteration at which the algorithm fails and returns $\bot$.

\begin{branchingrule}
	\label{br:1}
	For each $v \in \left( \bigcup_{i \in [\ibr-1]} V(P_i) \cup \bigcup_{j \in [\jbr-1]} V(Q_{\ibr,j}) \right) \setminus \bigcup_{i\in[k]} L_i$ create a new instance $(G, k, \ell, s, t, (L_i)_{i\in [k]})$ as follows: Graph $G$, integers $k$ and $\ell$ and vertices $s$ and $t$ remain unchanged. For the checkpoint lists, insert $v$ in $L_{\ibr}$ at index $\jbr+1$. Effectively, this causes the desired subpath $Q_{\ibr,\jbr}$ going from $u\rightarrow u'$ to split into two subpaths $u \rightarrow v$ and $v \rightarrow u'$.
\end{branchingrule}

\begin{lemma}
	Branching Rule \ref{br:1} is safe, that is, the input instance $(G,k,\ell,s,t,(L_i)_{i\in [k]})$ has a solution if and only if one of the output instances has a solution.
\end{lemma}

\begin{proof}
	Let $(P^\star_i)_{i\in [k]}$ be a solution to $I=(G,k,\ell,s,t,(L_i)_{i\in [k]})$. For each $P^\star_i$ for $i \in [k]$, let $(Q_{i,j}^\star)_{j \in [|L_i|-1]}$ be a list of subpaths connecting the terminals in $L_i$. By Lemma \ref{lemma:optimalUsesGreedyVertexSPPC2}, $Q_{\ibr,\jbr}^\star$ has non-empty intersection with the non-terminal vertices used in the greedy solution. 
	In one of the choices of $v$ tried by the Branching Rule, we thus have $v \in V(Q_{\ibr,\jbr}^\star)$. Thus, we may split the path $Q_{\ibr,\jbr}^\star$ into two subpaths $Q^{\star a}_{\ibr,\jbr}$ and $Q^{\star b}_{\ibr,\jbr}$  by separating it at vertex $v$. 
	Denote the new child instance with the split subpaths by $I' = (G,k,\ell,s,t,(L'_i)_{i\in [k]})$. 
	As stated in the branching rule, $L'_i = L_i$ for each $i \in [k]\setminus \{\ibr\}$. For $L_{\ibr}$, we have that $L'_{\ibr}[j] = L_{\ibr}[j]$ for each $j \in [\jbr]$. 
	Finally, we have $L'_{\ibr}[\jbr+1] = v$ and $L'_{\ibr}[j] = L_{\ibr}[j-1]$ for each $j \in [\jbr, |L_i|-1]$. 
	As the only thing that has changed is the insertion of a vertex into one list, we can easily see that we can use the constructed subpaths and set $Q'_{\ibr,\jbr} = Q^{\star a}_{\ibr,\jbr}$ and $Q'_{\ibr,\jbr+1} = Q^{\star b}_{\ibr,\jbr}$. 
	Since the paths $Q_{i,j}^\star$ are disjoint and each path $P^*_i$ fulfills the length constraints, the same holds for $Q'_{i,j}$ – therefore, we have constructed a solution to the instance $I'$.
	
	Conversely, assume that one of the instances constructed by Branching Rule \ref{br:1} has a solution $(P'_i)_{i\in [k]} = (Q'_{i,j})_{i \in [k], j \in [|L_i|]}$. Since $Q'_{\ibr,\jbr}$ and $Q'_{\ibr,\jbr+1}$ are disjoint paths going from $u \rightarrow v$ and from $v \rightarrow u'$, we can concatenate them resulting in a path $Q_{\ibr,\jbr}^\star$. Furthermore, we have to set $Q_{\ibr,j}^\star = Q'_{\ibr,j}$ for each $j \in [\jbr-1]$ and $Q_{\ibr,j}^\star = Q'_{\ibr,j+1}$ for each $j \in [\jbr+1,|L'_i|-2]$. Therefore, this results in a solution to the input instance.
\end{proof}

\begin{branchingrule}
	\label{br:2}
	For each $j' \leq \jbr$ and $v \in \left( \bigcup_{i \in [\ibr-1]} V(P_i) \cup \bigcup_{j \in [\jbr-1], j\neq j'} V(Q_{\ibr,j}) \right)$ create a new instance $(G, k, \ell, s,t, (L_i)_{i\in [k]})$ as follows: Graph $G$, integers $k$, $\ell$ remain unchanged. in $L$, insert $v$ in $L_{\ibr}$ at index $j'+1$. Effectively, this causes the desired subpath $Q_{\ibr,j'}$ going from $u\rightarrow u'$ being split into two subpaths $u \rightarrow v$ and $v \rightarrow u'$.
\end{branchingrule}

\begin{lemma}
	Branching Rule \ref{br:2} is safe, that is, the input instance $(G,k,\ell,s,t,(L_i)_{i\in [k]})$ has a solution if and only if one of the output instances has a solution.
\end{lemma}

\begin{proof}
	Let $(P^\star_i)_{i\in [k]}$ be a solution to $I=(G,k,\ell,s,t,(L_i)_{i\in [k]})$. For each $P^\star_i$ for $i \in [k]$, let $(Q_{i,j}^\star)_{j \in [|L_i|]}$ be a list of subpaths connecting the terminals in $L_i$. By Lemma \ref{lemma:optimalUsesGreedyVertexSPPC3}, some subpath $Q_{\ibr,j'}^\star$ has non-empty intersection with the non-terminal vertices used in the greedy solution. 
	In one of the choices of $v$ tried by the Branching Rule, we thus have $v \in V(Q_{\ibr,j'}^\star)$. The rest of the proof for this and the other direction is equivalent to the proof of Branching Rule 1.
\end{proof}

Since both branching rules are safe and in each step we increase the length of some terminal list, this algorithm will terminate and find a solution if there is one.

\subsection{Example Run}

\begin{figure}[t]
  \centering
  \includegraphics[width=0.6\textwidth, page=3]{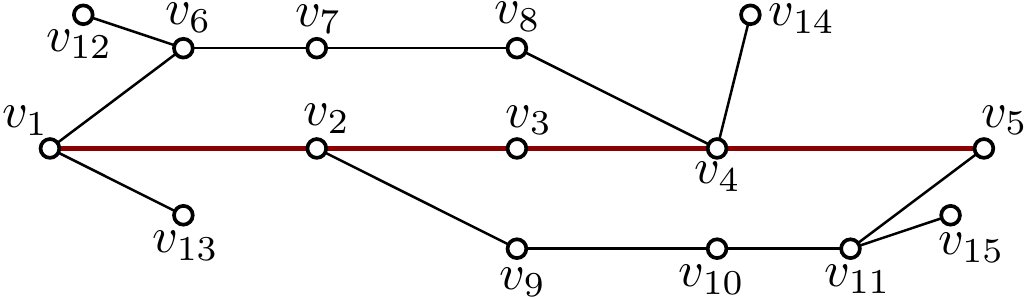} \\
  \caption{Graph $G$ from Figure \ref{fig:prel:exgr} and \ref{fig:theor:example} used as part of our instance. The first path $P_1$ found by GreedySPPC is marked in red.}
  \label{fig:theor:algo-run-example1}
\end{figure}

Finally, we want to present one example run of the algorithm on a simple instance.
Given the instance from Figure \ref{fig:theor:example} in Section \ref{ch:theor:prob-def}, we want to find $k=2$ disjoint paths of length at most $\ell=5$ going from $v_1$ to $v_5$. Therefore, we start with two checkpoint lists $L_1 = L_2 = \{v_1,v_5\}$. The algorithm starts by greedily determining the shortest $v_1$-$v_5$ path which is marked in red in Figure \ref{fig:theor:algo-run-example1}. 

\begin{figure}[t]
  \centering
  \includegraphics[width=0.6\textwidth, page=4]{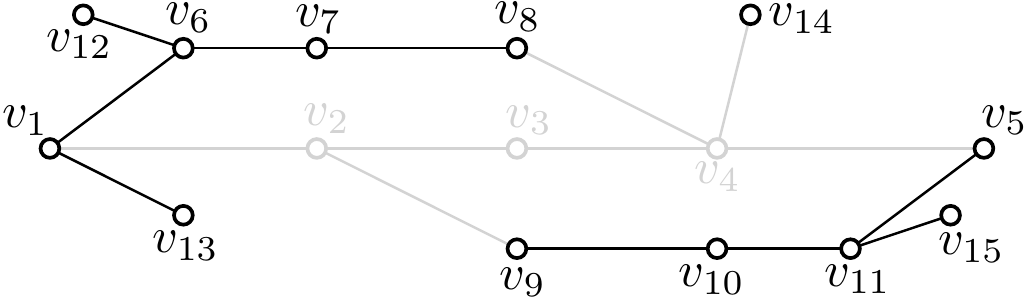} \\
  \caption{Working graph $G_2$ – After removal of the internal vertices of $P_1$, no $v_1$-$v_5$ path can be found.}
  \label{fig:theor:algo-run-example2}
\end{figure}

This gives us our first path $P_1$. Then, the internal vertices of the path get removed from the graph (see Figure \ref{fig:theor:algo-run-example2}), resulting in $v_1$ and $v_5$ being disconnected, thus the greedy algorithm failing in iteration $\ibr = 2$ because no $v_1$-$v_5$ path could be found. Because of this, we have to invoke Branching Rule 1 which means creating a new child instance for every internal vertex of the previously found paths: In our case those are the internal vertices of $P_1$, which are $v_2, v_3$ and $v_4$.

\begin{figure}[t]
  \centering
  \includegraphics[width=0.59\textwidth, page=5]{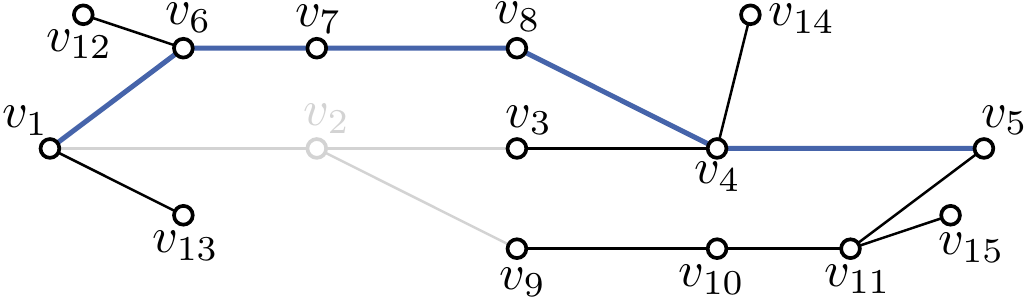} \\
  \caption{Working graph $G_{1,1}$ in the first child instance $I'$ – Vertex $v_2$ is a checkpoint in $L_2$, therefore it gets removed from this graph. Then, the greedy algorithm finds the path marked in blue as $P_1$.}
  \label{fig:theor:algo-run-example3}
\end{figure}

\begin{figure}[t]
  \centering
  \includegraphics[width=0.59\textwidth, page=6]{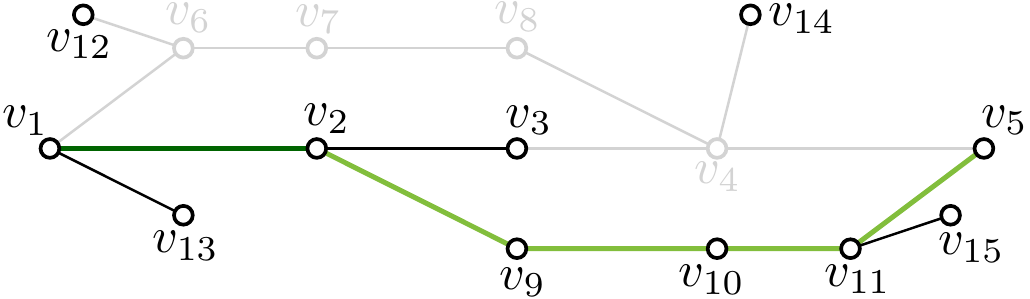} \\
  \caption{Working graph $G_{2}$ in the first child instance $I'$ – After $P_1$ was removed from the graph, finding the two subpaths  $v_1 \rightarrow v_2$ and $v_2 \rightarrow v_5$ marked in dark green and light green is straight forward.}
  \label{fig:theor:algo-run-example4}
\end{figure}

\begin{figure}[t]
  \centering
  \includegraphics[width=0.59\textwidth, page=2]{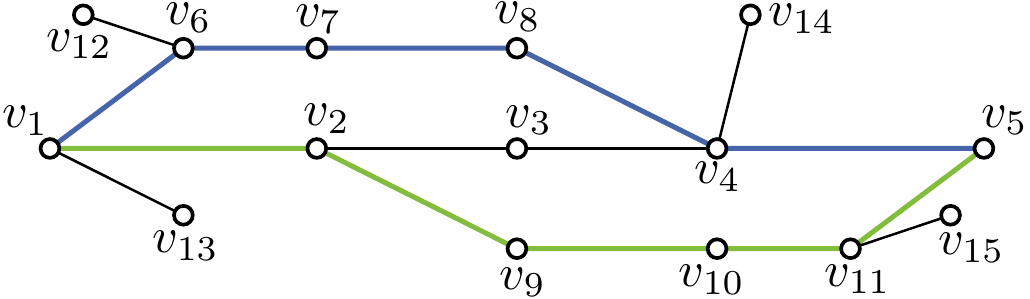} \\
  \caption{Final solution $\mathcal P = ((v_1, v_6, v_7, v_8, v_4, v_5), (v_1, v_2, v_9, v_{10}, v_{11}, v_{5}))$ computed by the algorithm.}
  \label{fig:theor:algo-run-example5}
\end{figure}

Our first child instance $I'$ therefore differs from its parent by the modified checkpoint list $L_2 = \{v_1, v_2, v_5\}$. This has the effect that as $v_2$ now is a checkpoint, it cannot be used by $P_1$ and therefore also is not part of the working graph $G_{1,1}$ (Figure \ref{fig:theor:algo-run-example3}). This causes the greedy algorithm to find the path marked in blue as $P_1$, which then gets removed from the graph resulting in our working graph $G_2$ (Figure \ref{fig:theor:algo-run-example4}). From here on, the subpaths $v_1 \rightarrow v_2$ and $v_2 \rightarrow v_5$ can be easily found, thus giving us a solution (Figure \ref{fig:theor:algo-run-example5}).

\subsection{Runtime Complexity}
\label{ch:theor:runtime}

We can bound the running time of the algorithm by considering bounds on the number of child instances produced. 

All of the paths together cannot have more than $k\cdot \ell$ vertices in total. As we branch over previously found vertices and thus over a subset of a possible solution, $k \cdot \ell$ also is an upper bound for the number of child instances produced while applying Branching Rule \ref{br:1}. In Branching Rule \ref{br:2}, however, we have an additional factor of up to $\ell$: As by Observation \ref{obs:theor:trivial}, any terminal list having more than $\ell + 1$ elements make the instance infeasible. Therefore, the number of positions we need to additionally branch over is bounded by $\ell$.

At the same time, we need to bound the number of levels that can exist in our search tree. As in each step one more vertex gets added to some terminal list, the number of ``free'' vertices decreases by 1 in each level. By the previous argument we can have at most $k \cdot \ell$ vertices in our paths in total, therefore we can also use this to bound the number of levels. 

Lastly, as we are assuming graphs without edge weights, shortest path computations for the greedy part are feasible in runtime that is linear with regards to the number of vertices and edges of the graph by using breadth first search, which results in the following theorem:

\begin{theorem}
	\SPP can be solved in $(k\cdot \ell^2)^{k\cdot \ell} \cdot O(n+m)$ time by using a search-tree algorithm after transforming it to an instance of \SPP \textsc{with Checkpoints}.
\end{theorem}

%% file: chapters/chapter-heuristic-improvements.tex
\newcommand{\bigcupdot}{\ \cdot \hspace{-7.5pt}\bigcup}
\newcommand{\ibr}{{i_\beta}}
\newcommand{\jbr}{{j_\beta}}

\onlyinsubfile{\setcounter{chapter}{2} \newcommand{\figpath}{../figures/}}
\chapter{Heuristic Improvements}
\label{ch:heur}
While Chapter \ref{ch:theor} sets the theoretical foundation on which the algorithm is based, the performance of a search-tree algorithm highly depends on the specific heuristic optimizations used. In this chapter we introduce multiple such optimizations, which we divide in multiple categories: First, we show ways to decrease the size of the graph in advance and some initial heuristics which allow for detecting trivial instances beforehand. Then, we present techniques to detect negative instances in the search tree early on, followed by improvements to the greedy phase. Furthermore, we show another type of improvement which is used to break symmetries in the search tree. Finally, we look into optimizing the order of subproblem expansion for faster discovery of solutions, if they exist.

\section{Preprocessing the Input Graph}
\label{ch:heur:preproc}

Our input graphs in general might be arbitrarily large, however, for computing multiple disjoint $s$-$t$ paths of bounded length, only local parts of the graph around $s$ and $t$ are relevant. Therefore, we can reduce the size of our working graph using the following lemmas: 

\begin{lemma}
	Let $N_\ell(s)$ and $N_\ell(t)$ be the $\ell$-neighborhood of $s$ resp. $t$. For any solution to an instance of SPP(C), all solution paths lie within the intersection $N_\ell(s) \cap N_\ell(t)$ between those two sets.
	\begin{proof}
	\vspace{-1.2em}
		Assume that a solution $S$ used a vertex $v$ that is, without loss of generality, not part of $N_\ell(s)$. This means that as the distance from $s$ to $v$ is larger than $\ell$, the distance from $s$ to $t$ also must be larger than $\ell$. That is a contradiction to $S$ being a solution.
	\end{proof}
\end{lemma}

\begin{lemma}
	Let $N_{\lfloor\ell/2\rfloor}(s)$ and $N_{\lfloor\ell/2\rfloor}(t)$ be the $\lfloor\ell/2\rfloor$-neighborhood of $s$ resp. $t$. For any solution to an instance of SPP(C), all solution paths lie within the union $N_{\lfloor\ell/2\rfloor}(s) \cup N_{\lfloor\ell/2\rfloor}(t)$ between those two sets.
	\begin{proof}
		Assume that a solution $S$ used a vertex $v$ that is neither part of $N_{\lfloor\ell/2\rfloor}(s)$ nor part of $N_{\lfloor\ell/2\rfloor}(t)$. This means that the distances from $s$ to $v$ and from $v$ to $t$ are both larger than $\lfloor\ell/2\rfloor$. If $\ell$ was even, this gives us a total length of $\ell + 2$, if it was odd, a length of $\ell + 1$ – both a contradiction to $S$ being a solution.
	\end{proof}
\end{lemma}

Putting those two lemmas together, we get the set $V' = N_\ell(s) \cap N_\ell(t) \cap (N_{\lfloor\ell/2\rfloor}(s) \cup N_{\lfloor\ell/2\rfloor}(t))$ and from here on we can work with the graph $G[V']$, thus decreasing the runtime of all the polynomial algorithms involved in the different heuristics and the greedy algorithm itself.

However, for increasing values of $\ell$ one has to recognize that the potential of any relevant reduction of size decreases, as many real-world and/or randomized graphs have rather small diameter, thus making all of the vertices of the graph reachable within $\ell$ or even $\lfloor\ell/2\rfloor$ steps. The popular ``six degrees of separation'' \cite{Barabasi2002} in social networks also give an intuitive explanation why for many graphs a higher choice of $\ell$ often does not allow for effective preprocessing.

\section{Detecting Trivial Instances}

Before we even start traversing the search tree, we can make use of some simple initial heuristics to detect trivial instances.

\subsubsection{Minimum $s$-$t$ separator}
\label{ch:heur:prealg:minsep}

A \textit{minimum $s$-$t$ vertex separator} (sometimes called vertex cut) is a minimal set of vertices that upon removal disconnect $s$ and $t$ into separate connected components. By Menger's Theorem \cite{Menger1927} we know that the size of such a minimum $s$-$t$ separator equals the maximum number of vertex-disjoint $s$-$t$ paths. So clearly this gives an upper bound for the number of $\ell$-bounded vertex-disjoint $s$-$t$ paths. Therefore, by computing a minimum $s$-$t$ vertex separator for our graph, which is possible in polynomial time \cite{FordFulkerson1956}, we know that no solution can exist, if its size is smaller than $k$.

\subsubsection{Vertex-disjoint $s$-$t$ paths of minimum total length}
\label{ch:heur:prealg:dpmintotal}

The problem of computing $k$ $s$-$t$ paths of minimum total length is solvable in polynomial time by using, e.g., an adaption of Suurballes algorithm \cite{Suurballe1974}. Therefore, we can use this problem both to detect trivial \SPP yes- and no-instances based on a minimal solution: 

If the longest path of a collection of $k$ paths of minimum total length is shorter than $\ell$, then the solution directly translates to a solution of our \SPp-instance: Each of the $k$ $s$-$t$ paths have length at most $\ell$ and they are vertex disjoint, thus it is a yes-instance. 

On the other hand, if the total length of the paths in a collection of $k$ paths with minimum total length is larger than $k \cdot \ell$, then we know that our \SPp-instance is a no-instance: If a solution to \SPP existed, its total length would definitely be less than $k \cdot \ell$. However, this solution would also qualify as a solution to the minimum total length problem, thus violating the minimality of the solution found by Suurballes algorithm.

\section{Pruning the Search Tree}

While we showed the theoretical bounds for the size of the search tree in Chapter \ref{ch:theor:runtime}, in practice we do not even want to get close to fully expanding the search tree. By the nature of the problem, there exist many situations where we know early on that expanding a search node can never yield a solution. The following heuristics help in recognizing such situations – in parentheses we introduce short codes for identifying heuristics later on. 

\subsubsection{\underline{B}reak by \underline{c}heck\underline{p}oint list \underline{l}ength (\texttt{b-cpl})}
\label{h:b-cpl}

For the theoretical foundation in Chapter \ref{ch:theor}, we use the trivial break condition that any instance containing a checkpoint list of length longer than $\ell$ cannot have a solution. In practice, however, in most cases this means filling checkpoint lists with vertices that are not even close to each other and therefore unnecessarily blowing up the size of the search tree. To avoid this, one can further refine the breaking mechanism using the following lemma:

\begin{lemma}
	Given an instance $(G, k, \ell, s, t, (L_i)_{i\in [k]})$, if there is a checkpoint list $L_i$ that has more than $\lfloor\ell/2+1\rfloor$ elements but contains less than $2(|L_i|-1)-\ell$ pairs of consecutive elements that are adjacent in $G$, then no solution can exist.

	\begin{proof}
		In general, if $L_i$ does not contain any pair of consecutive adjacent elements, then any $L_i$-path has a length of at least $2(|L_i|-1)$. Each pair of consecutive elements which are adjacent brings down this length by 1. Therefore, if $s$ denotes the number of pairs of consecutive adjacent elements in $L_i$, then any $L_i$-path has a length of at least $2(|L_i|-1)-s$.

		Now, assume that a checkpoint list has more than $\lfloor\ell/2+1\rfloor$ elements, but only $2(|L_i|-1)-\ell-1$ pairs of consecutive adjacent elements. This means, that any $L_i$-path has a length of at least $2(|L_i|-1)-(2(|L_i|-1)-\ell-1) = \ell + 1$. Therefore, no solution satisfying the checkpoint list $L_i$ can exist.
	\end{proof}
\end{lemma}

\subsubsection{\underline{B}reak by \underline{s}hortest \underline{p}aths between consecutive checkpoint pairs (\texttt{b-sp})}
\label{h:b-sp}
A stricter, but computationally more expensive variant of the previous heuristic is the following:

\begin{lemma}
	Given an instance $(G, k, \ell, s, t, (L_i)_{i\in [k]})$, if for any checkpoint list $L_i$ the sum of the lengths of the shortest paths in $G$ between consecutive checkpoints exceeds $\ell$, then no solution to the instance can exist.
	\begin{proof}
		Determining all shortest paths between consecutive pairs of checkpoints in $L_i$ is a lower bound for the length of an $L_i$-path. Clearly, if the lower bound of $L_i$-path exceeds $\ell$, then no $L_i$-path with length at most $\ell$ can exist.
	\end{proof}

\end{lemma}

\section{Improving the Greedy Phase}
\label{ch:heur:improv-greedy}

Another factor that controls the size of the search tree is the number of subpaths determined by the greedy algorithm before it fails. This controls the branching factor as for each vertex in an already computed subpath, a new instance is created. Therefore, it is of interest to detect as early as possible that a greedy run cannot reach a solution. We try to achieve this by using the following additional failure detection mechanism which runs each time when an additional complete shortest $s$-$t$ path was determined.

\subsubsection{\underline{D}etect failure by $s$-$t$ \underline{m}inimum \underline{s}eparator in working graph (\texttt{d-ms})}
\label{h:d-ms}
Exactly like in the pre-search-tree heuristics in Section \ref{ch:heur:prealg:minsep} we can calculate a minimum $s$-$t$ separator in the working graph $G_i$, in which only the vertices of the already found paths were removed. If the goal is to find $k$ disjoint paths and after $i$ iterations, $i$ paths were already found, then in the working graph $G_i$ the size of the minimum $s$-$t$-separator has to be at least $k-i$. If that would not be the case, this would mean that in the graph $G_i$, no $k-i$ vertex-disjoint $s$-$t$ paths existed which implies that from this situation a solution can never be found.

\subsubsection{Dealing with \texttt{d-ms}}

While the above descriptions tell us when continuing a greedy run is pointless, this does not tell us what exactly to do about it. Naively, one could assume that we can just invoke Branching Rule 1 (or 2), but of course this has to be justified as the failure conditions, as we called them in Chapter \ref{ch:theor}, are completely different here. And the doubt is reasonable: If we recall Failure Condition 1 and 2, there we know that the cause can be limited to a specific subpath or at least a specific set of subpaths. In this situation, where in iteration $\ibr$ of the greedy run, \texttt{d-ms} fires, the cause cannot be limited in such a way. Let us call this situation from now on Failure Condition 3. We only know that at the beginning of iteration $\ibr$, which is the iteration of the greedy run where the execution fails, the remaining graph has a property which makes the further search pointless. This, however, can not be used to argue that path $P_\ibr$ has to use some previously used vertex – the checkpoint list $L_\ibr$ could in fact be perfectly fine and it could, e.g., be some position in list $L_{\ibr+2}$ which has to use a previously used vertex.
The key point is, that we only know that \textit{some} subpath of \textit{all} paths remaining to be computed has to use some of the previously used vertices, which in consequence means that none of the existing branching rules are applicable to this situation.

To formulate a new branching rule, we therefore need to state a new lemma for this situation:

\begin{lemma}
	\label{lemma:optimalUsesGreedyVertexSPPC4}
	Given an SPPC instance $I = (G, k, \ell, s, t, (L_i)_{i\in [k]})$, assume Algorithm \ref{alg:greedySPPC} together with the improvement \texttt{d-ms} failed in the beginning of iteration $\ibr$ of the outer loop because of Failure Condition 3. Let $\mathcal P$ be the returned collection of $\ibr-1$ disjoint paths $(P_i)_{i \in [\ibr-1]}$. If $I$ is a yes-instance and thus a solution $(P^*_i)_{i \in [k]}$, each consisting of subpaths $(Q^*_{i,j})_{j\in[|L_i|-1]}$ exists, then there exists some subpath $Q^*_{i',j'} \in (Q^*_{i,j})_{i \in [\ibr, k], j \in [|L_i|-1]]}$ which uses some non-terminal vertex of the greedily computed collection of paths $\mathcal P$. In formulas,
	\[ \bigg(  \bigcup_{i \in [\ibr,k]} \bigcup_{j \in [|L_i|-1]} V(Q^*_{i,j}) \bigg) \cap \bigg( \bigcup_{i \in [\ibr-1]} V(P_i) \setminus \bigcup_{i\in[k]} L_i \bigg) \Bigg) \neq \emptyset.\]
	
\begin{proof}
	Assume, the intersection was empty. 
	This means that any solution subpath $Q^*_{i,j}$ for $i \in [\ibr, k], j \in [|L_i|-1]$ does not contain an internal vertex of any other path $P_i$ for $i \in [\ibr-1]$. This means, that each subpath $Q^*_{\ibr,j}$ for each $j \in [\jbr]$ must be contained in the graph $G_{\ibr}$. This means, that $k-(\ibr-1)$ $s$-$t$ paths of length at most $\ell$ still exist in the graph $G_{\ibr}$. This is a contradiction to Failure Condition 3 occurring. Therefore, the intersection cannot be empty.
	\end{proof}
\end{lemma}

This lemma now allows us to introduce the new branching rule:
\begin{branchingrule}
	\label{br:3}
	For each $i \in [\ibr, k]$, each $j \in [|L_i|-1]$ and each of the vertices $v \in \bigcup_{i \in [\ibr-1]} V(P_i) \setminus L_i$ create a new instance $(G, k, \ell, s,t, (L_i)_{i\in [k]})$ as follows: Graph $G$, integers $k$, $\ell$ remain unchanged. For the checkpoints lists, insert $v$ in $L_{\ibr}$ at index $j'+1$. Effectively, this causes the desired subpath $Q_{\ibr,j'}$ going from $u\rightarrow u'$ being split into two subpaths $u \rightarrow v$ and $v \rightarrow u'$.
\end{branchingrule}

\begin{lemma}
	Branching Rule \ref{br:3} is safe, that is, the input instance $(G,k,\ell,s,t,(L_i)_{i\in [k]})$ has a solution if and only if one of the output instances has a solution.
\end{lemma}

\begin{proof}[Proof sketch]
	Using Lemma \ref{lemma:optimalUsesGreedyVertexSPPC4}, the proof follows the exact same pattern as the proofs for the safeness of Branching Rule 1 and 2.
\end{proof}

As a last remark one has to consider that as soon as this branching rule is employed, the worst-case branching factor worsens by a factor of $k$, resulting in a new worst-case asymptotic runtime of $(k^2\cdot\ell^2)^{k\cdot\ell}\cdot n^{O(1)}$. Therefore, the initial goal of reducing the branching factor is not really improved by this approach and it is unclear, whether it can even be considered an improvement. The experiments in Chapter \ref{ch:exper} will give insights on this.

\section{Breaking Symmetries}
\label{ch:heur:symmetry-breaking}

\begin{figure}[t]
\centering
  \includegraphics[width=0.8\textwidth]{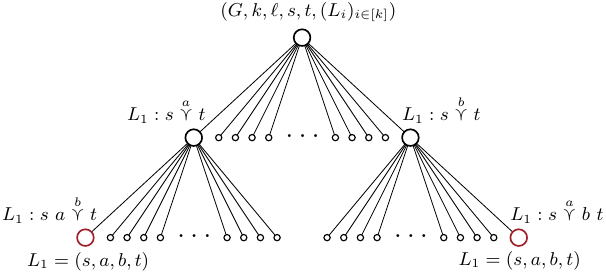}
  \caption{A simplified example for how symmetries in our search tree can emerge.}
  \label{fig:heur:symmetrie-ex}
\end{figure}

By the nature of our branching strategy, symmetries in the search tree can easily occur. As depicted in Figure \ref{fig:heur:symmetrie-ex}, a checkpoint list containing e.g. $(s,a,b,t)$ can be a result of first branching with $a$ and then with $b$ or conversely. Clearly, we want to avoid such situations, as it means recomputing things that already were computed.

\subsubsection{\underline{B}reak using \underline{f}orbidden \underline{i}ntervals (\texttt{b-fi})}
\label{h:b-fi}
\newcommand{\forbidint}[3]{$(#1 \stackrel{#3}{\cancel{\curlyvee}} #2)$}
 
A possible solution to this exploits the following fact: When looking at an instance $I$ of the search tree and a branching rule is used, the generated child instances differ in just one additional vertex inserted in one of the checkpoint lists. 

Now consider that in the first child instance $I'_1$, vertex $x$ gets inserted in checkpoint list $L_i$ between vertices $a$ and $b$. Then, assume that the attempt of solving $I'_1$ returns without success. From this we can deduce that any solution to the parent instance $I$ can never contain a path visiting $a$, $x$ and $b$ in this order – otherwise, it would have been found in the process of trying to solve $I'_1$. Therefore, we call \forbidint{a}{b}{x} a \textit{forbidden interval} for $x$.

\begin{figure}[t]
\centering
  \includegraphics{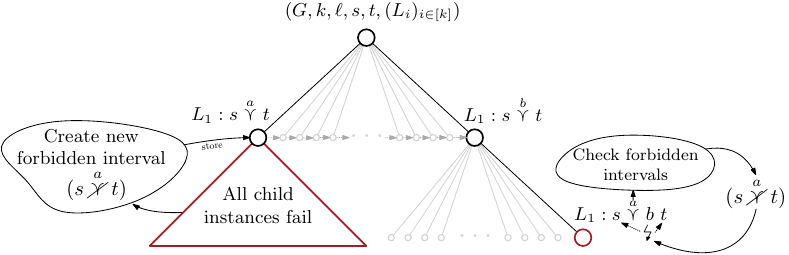}
  \caption{Basic principle of the forbidden interval heuristic}
  \label{fig:heur:forbidint-ex}
\end{figure}

This means, that for any of the following child instances $I'_2, I'_3, ...$, at any point of further application of the branching rule, vertex $x$ can never be inserted between $a$ and $b$ in checkpoint list $L_i$, as due to above reasoning such a configuration could never yield a solution. Figure \ref{fig:heur:forbidint-ex} illustrates this situation.

Because a forbidden interval can never occur in any of the solutions to another child instance, we can even go one step further and modify our definition of the working graph $G_{i,j}$ in Algorithm \ref{alg:greedySPPC2} to remove all vertices $v$ from the graph, that are part of a forbidden interval \forbidint{a}{b}{v} such that $a$ occurs in checkpoint list $L_i$ before index $j$ and $b$ occurs after index $j$.

\subsubsection{Cache instances}

Another approach that we investigated is storing all previously solved instances in some data structure. However, it was quickly clear that space usage and lookup times are not feasible with simple caching solutions because we have an exponential number of possible subproblems. Nevertheless, this could be an interesting focus for future work.

\section{Choosing Promising Subproblems}

The branching rules introduced in Chapter \ref{ch:theor} specify the set of vertices to branch over, but not the order in which the subproblems are processed. Therefore, good heuristics for choosing the right subproblem can greatly decrease the time necessary to find a solution – for negative instances, however, they can not yield any improvement, as the order of subproblems does not matter in negative cases.

\subsubsection{\underline{C}hoose by lowest \underline{dist}ance (\texttt{c-dist})}
\label{h:c-dist}
Let $BS$ denote the branching set, namely the vertices $v$ specified in Branching Rule \ref{br:1}/\ref{br:2}/(3), which get inserted into an instance at a specified position between vertices $u$ and $u'$ to yield a child instance.

We can now sort the vertices in $BS$ by the sum of the distance $d(u,v)$ and $d(u',v)$ in the full input graph. The motivation behind this is that good choices for checkpoints between $u$ and $u'$ are those which are relatively close to both.

\subsubsection{\underline{C}hoose by \underline{p}ath\underline{l}ength at point of failure (only for BR2, \texttt{c-pl})}
\label{h:c-pl}
In Branching Rule \ref{br:2}, on top of choosing the order of the vertices to be inserted, there is an additional degree of freedom choosing the position of insertion within the checkpoint list. This stems from the fact that Branching Rule \ref{br:2} gets invoked in the case of a path getting too long and therefore not only the last, but any subpath of the failed path possibly being the source of failure. 

A potential strategy here is to choose positions corresponding to longer subpaths. The naive motivation is  that, while the shorter paths are already short, the long paths might have more potential to achieve shorter total path length.

%% file: chapters/chapter-implementation-details.tex
\newcommand{\bigcupdot}{\ \cdot \hspace{-7.5pt}\bigcup}
\newcommand{\ibr}{{i_\beta}}
\newcommand{\jbr}{{j_\beta}}
\newcommand{\forbidint}[3]{$(#1 \stackrel{#3}{\cancel{\curlyvee}} #2)$}

\onlyinsubfile{\setcounter{chapter}{2} \newcommand{\figpath}{../figures/}}
\chapter{Implementation Details}
\label{ch:impl}

We implemented the theoretical algorithm from Chapter \ref{ch:theor} together with the improvements from Chapter \ref{ch:heur} to be able to evaluate the different improvements and the overall performance of the algorithm. In this chapter, we describe how certain parts of the algorithm are actually implemented. We start by giving some general implementation details in Section \ref{ch:impl:gener}, go on by describing the main skeleton for the search-tree algorithm in Section \ref{ch:impl:st} and finally in Section \ref{ch:impl:improv} go into detail about the implementation of several of the improvements introduced in Chapter \ref{ch:heur}.

\section{General Implementation Details}
\label{ch:impl:gener}

The implementation was written in Java 17 \cite{JDK} and uses graph algorithms and data structures from the \textit{JGraphT} library \cite{MichailEtAl2020}, which is one of the most-used Java graph libraries. The source code of our implementation is published online \cite{Huber2022}. As \SPP concerns simple, undirected, unweighted, loopless graphs, the data structure of choice is the class \texttt{SimpleGraph<V,E>}. Being generic, in place of \texttt{V} and \texttt{E} it requires a vertex and edge class to be specified. For this thesis the vertices only need to be identifiable by an integer id, which is implemented in a custom \texttt{Vertex} class. There also is no necessity to store data on the edges, therefore the given JGraphT class \texttt{DefaultEdge} was used.

Some computations like calculating the shortest paths are used many times throughout the algorithm. Therefore we address the way such general things were solved before specifying details of certain steps in the algorithm.
\begin{description}
	\item[Dynamically modifying the graph.] JGraphT comes with a useful \texttt{MaskSubgraph} class, which allows to dynamically access the subgraph induced by a vertex set based on a filtering function. This is necessary in multiple steps of the algorithm: For calculating shortest paths between checkpoints we never want to use other checkpoints, therefore we filter those. Also, as soon as we have greedily determined a shortest path we need to exclude the vertices of that path from the ongoing search.
	
	\item[Calculating shortest paths.] In general, \texttt{BFSShortestPath}, a JGraphT class, is being utilized, allowing shortest paths to be computed in $O(n+m)$ time. We use it in the shortest path checkpoint heuristic and for the general greedy algorithm itself. For calculating shortest paths between checkpoints for sorting them by their distance (\texttt{c-dist}), as this is such a frequent operation, we additionally store the computed path lengths in a map to be able to access them efficiently.
	
\end{description}

\section{Search-Tree Algorithm}
\label{ch:impl:st}

 The main entry point is the class \texttt{SPPAlgorithm} and its method \texttt{Solution} \texttt{run( Instance i)}. It takes an instance, consisting of a \texttt{Graph<V,E>}, an integer \texttt{ell}, vertices \texttt{s} and \texttt{t} and a list of \texttt{Checkpoints} objects, which themselves are just lists of \texttt{Vertex} objects. The value of $k$ is implicitly given by the number of such checkpoint lists. Furthermore, the instance contains a list of \texttt{ForbiddenInterval} objects, which is used if the improvement \texttt{b-fi} is enabled and each just contains three \texttt{Vertex} objects, for storing the start, end and the affected vertex for the forbidden interval.
 
 Depending on if they are enabled, first the preprocessing and the trivial instance recognition is run – their implementation details  are described in the next section. Then, the actual search-tree algorithm starts and its traversal is realized in a recursive fashion. For this, the method \texttt{Result solve(Instance i)} in the class \texttt{SPPAlgorithm} is called, which is almost exactly an implementation of the pseudocode in Algorithm \ref{alg:searchTreeSPPC}. There, it is first checked, whether the instance is a trivial no-instance, recognized by either our naive mechanism checking for a checkpoint list length of more than $\ell+1$ or by one of our improvements \texttt{b-cpl} or \texttt{b-sp}. If it is not a trivial no-instance, we try solving the instance greedily. To do so, we call the method \texttt{Result} \texttt{trySolveGreedy(Instance i)}, which is again mostly an implementation of the pseudocode given in Algorithm \ref{alg:greedySPPC2}. It returns a \texttt{Result} object, containing a complete solution in case of success, or a partial solution together with the corresponding failure condition in case of failure. Based on that condition, one of the 2 (or, if \texttt{d-ms} is enabled, 3) branching rules is invoked. Within the branching rule, the branching set is first ordered by the corresponding active heuristic (\texttt{c-dist}, \texttt{c-pl}) and then recursive calls to \texttt{solve} are made in the corresponding order. As soon as one of the recursive calls returns with success, we return the solution – otherwise, we recurse over all branches before returning with failure.

\section{Heuristic Improvements}
\label{ch:impl:improv}

While some of the improvements, like ordering paths by their length or calculating shortest paths between checkpoints, are mostly straight-forward and their implementation was already briefly touched in Section \ref{ch:impl:gener}, we want to go into detail about some of the improvements where the specific implementation was not yet discussed.

\subsection{Preprocessing}

The preprocessing routines described in Chapter \ref{ch:heur:preproc} allow us to delete irrelevant, non-reachable vertices from the graph. This is done using breadth-first search, for which JGraphT provides a class \texttt{BreadthFirstIterator} which allows us to traverse the graph and at the same time keep track of the depth (meaning the distance from the start vertex) of the currently visited vertex. We stop the search as soon as we reach a depth of larger than our designated maximum radius. Additionally, we also get rid of degree-1-vertices as they can never be an internal vertex of any path.

\subsection{Detecting Trivial Instances}
\label{ch:impl:improv:trivial}

To detect one type of trivial instances, we want to determine a minimum $s$-$t$ separator (see Chapter \ref{ch:heur:prealg:minsep}). JGraphT unfortunately does not offer a built-in algorithm to compute such a separator, but it does come with a class \texttt{EdmondsKarpMFImpl} which using the method \texttt{calculateMinCut} allows us to calculate the minimum $s$-$t$ edge cut in a (directed) graph in running time $O(nm^2)$. 

A similar situation arises regarding the computation of $k$ internally vertex-disjoint $s$-$t$ paths of minimal total length (see Chapter \ref{ch:heur:prealg:dpmintotal}). In JGraphT, only a variant of Suurballes algorithm for finding $k$ \textit{edge}-disjoint $s$-$t$ paths of minimal total length in a directed graph is implemented. 

\begin{figure}[t]
\centering
  \includegraphics[width=0.3\textwidth]{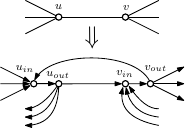}
  \caption{Example for the effect of Transformation \ref{transf:undirtodir} to two adjacent vertices}
  \label{fig:heur:transformation-expl}
\end{figure}

Therefore, to still be able to use those algorithms to get results wrt. to vertices and not to edges, we need to make use of the following commonly known transformation:

\begin{transformation}
	\label{transf:undirtodir}
	Let $G = (V,E)$ be any undirected graph. Construct a directed graph $G' = (V',E')$ with \begin{gather*}V' = \{ v_{in}, v_{out} \mid v \in V\} \text{ and} \\ E' = \{(v_{in},v_{out}) \mid v \in V\} \cup \{(u_{out},v_{in}), (v_{out},u_{in}) \mid \{u,v\} \in E\}.\end{gather*}
	In natural language, this means replacing each vertex by two new vertices, an ``in''-vertex and an ``out''-vertex, which are connected by a directed arc. Every (undirected) edge then gets replaced by two directed arcs, each connecting the out-vertices to the in-vertices of the adjacent vertices. A path from $s$ to $t$ of length $\ell$ in the old graph is now a path from $s_{out}$ to $t_{in}$ with length $2\ell-1$. Figure \ref{fig:heur:transformation-expl} shows an example.
\end{transformation}

\begin{lemma}
	\label{lemma:transf-mincut}
	For any graph $G$, a minimum edge cut $C$ of graph $G'$ obtained by Transformation \ref{transf:undirtodir} has the same size as a minimum vertex separator $S$ in the original graph~$G$. 
	\begin{proof}[Proof sketch]
		We can safely assume that $C$ only contains ``internal'' arcs of the type $(v_{in},v_{out})$. This is the case because for each arc of the type $(v_{out},u_{in})$, we can ``charge'' it to the arc $(v_{in},v_{out})$ or $(u_{in},u_{out})$ – in our case it would not make sense that two consecutive edges incident to the same vertex are in a minimum edge cut, thus neither $(v_{in},v_{out})$ nor $(u_{in},u_{out})$ are part of $C$. Our minimum edge cut $C$ now directly transforms to a minimum vertex separator $S$ by choosing for each ``internal'' arc $(v_{in},v_{out}) \in C$ of $G'$ the vertex $v$ of $G$ to be part of~$S$. Similarly, a minimum vertex separator $S'$ for $G$ directly transforms to a minimum edge cut $C'$ in $G'$ by just choosing the ``internal'' edges associated with each of the vertices in $S'$.
		\renewcommand{\qedsymbol}{}
	\end{proof}
\end{lemma}

\begin{lemma}
	\label{lemma:transf-dpmintotal}
	For any graph $G$, the collection $\mathcal P'$ of $k$ edge-disjoint $s_{out}$-$t_{in}$ paths of minimum total length in the graph $G'$ obtained by Transformation \ref{transf:undirtodir} directly translates to a set $\mathcal P$ of $k$ vertex-disjoint $s$-$t$ paths in $G$ of minimum total length. If the total length of $\mathcal P'$ in $G'$ is $\ell'$, then the total length of $\mathcal P$ in $G$ is $\ell = (\ell'+k)/2$.
	
	\begin{proof}[Proof sketch]
		It is clear that every $s$-$t$ path of length $\ell$ in $G$ corresponds to an $s_{out}$-$t_{in}$ path of length $2\ell-1$ in $G'$. An $s_{out}$-$t_{in}$ path of length $\ell'$ in $G'$ therefore corresponds to an $s$-$t$ path of length $(\ell'+1)/2$ (note, that $\ell'$ is always odd). Furthermore, any collection of edge-disjoint $s_{out}$-$t_{in}$ paths in $G'$ corresponds to a set of vertex-disjoint $s$-$t$ paths in $G$, as the single arcs going from $v_{in}$ to $v_{out}$ prohibit the usage of a vertex on multiple paths. Together, this results in $k$ paths in $G'$, each of length $\ell'_i$ for $i \in [k]$, directly corresponding to $k$ paths in $G$ of length $(\ell'_i+1)/2$. Therefore, the total length of all paths in $G$ amount to $\sum_{i \in [k]} ((\ell'_i+1)/2) = (\sum_{i \in [k]} \ell'_i + k)/2 = (\ell' + k)/2$. As this chain of arguments works in both directions, one can therefore derive that the optimality of the former edge-disjoint case implies the optimality of the vertex-disjoint case derived from it.
			\renewcommand{\qedsymbol}{}
	\end{proof}
\end{lemma}

For calculating a minimum vertex separator, we therefore apply this transformation and then calculate a minimum $s_{out}$-$t_{in}$ edge cut $C$ on $G'$ and make use of Lemma \ref{lemma:transf-mincut}. To compute a collection of $k$ vertex-disjoint $s$-$t$ paths of minimum total length, Lemma \ref{lemma:transf-dpmintotal} allows us to calculate a set of minimum total-length edge-disjoint paths on the transformed graph.

\subsection{Symmetry Breaking using Forbidden Intervals}

In the beginning, we start with an empty forbidden interval list. As described in Section \ref{ch:heur:symmetry-breaking}, if a child instance created by inserting $v$ between $u$ and $u'$ returns without success, then we append a new forbidden interval \forbidint{u}{u'}{v} to the list. The updated list always gets passed to new child instances and in the working graph we always mask the vertices belonging to ``active'' forbidden intervals, meaning that the checkpoint list contains the vertices $u$ and $u'$ in that order and the subpath currently being computed lies somewhere between those two vertices. 

\subsection{Additional Greedy Failure Condition}

The greedy improvement \texttt{d-ms}, which checks for the minimum vertex separator in the working graph after each iteration of the greedy algorithm, is basically the same as the trivial instance recognition mechanism described in Chapter \ref{ch:heur:prealg:dpmintotal} and Section \ref{ch:impl:improv:trivial}. Therefore, the implementation uses the same Transformation \ref{transf:undirtodir} as used for the trivial instance recognition.

%% file: chapters/chapter-experimental-results.tex
\onlyinsubfile{\setcounter{chapter}{2} \newcommand{\figpath}{../figures/}}
\addtocontents{toc}{\protect\newpage}
\chapter{Experimental Results}
\label{ch:exper}

To be able to judge the performance of the several improvements and the algorithm in general, we perform an experimental evaluation. The results are presented in this chapter. In Section \ref{ch:exper:effectiveness}, we compare different configurations of the improvements and check their effectiveness. Then, in Section \ref{ch:exper:performance}, we measure the general performance of the algorithm for different types of graphs and different values of $k$ and $\ell$ and conclude with a discussion of the results in Section \ref{ch:exper:dicuss}. 

\section{Technical Setup and Dataset}
\label{ch:exper:setup-dataset}

The experiments were run on a computing cluster with 16 nodes, each having 160GB RAM and two 10-core Intel Xeon E5-2640 v4, 2.40GHz. It runs a Linux system with kernel version 4.15.0. The program itself is run using Java 17. The results were then analyzed and visualized in Python 3.10.4 using the libraries pandas (1.5.0), numpy (1.23.3) and seaborn (0.12.0).

The dataset we use in the experiments was gathered by Nadara et al.\ \cite{NadaraEtAl2018} for conducting experiments on computing coloring numbers \cite{NadaraEtAl2019}. It consists of a total number of 92 sparse graphs taken from diverse sources which were then partitioned into categories \texttt{small}, \texttt{medium}, \texttt{big} and \texttt{huge} based on their size. In our experiments, for succinctness, we use the first three categories and discard the \texttt{huge}-group, leaving us with 77 graphs. The graphs are taken from real-world data such as, e.g., social networks, gene expressions or infrastructure data, as well as random planar graphs and random graphs with bounded expansion. Table \ref{t:ex-statistics} gives an overview on the dataset.

\begin{table}[t]
\caption{Statistics of test groups given by Nadara et al.\ \cite{NadaraEtAl2019}}
\begin{tabular}{lrrrrrrrr}
\toprule & \multicolumn{4}{c}{$|V(G)|$} & \multicolumn{4}{c}{$|E(G)|$} \\
\cmidrule(l){2-5} \cmidrule(l){6-9} group & min & med & avg & $\max$ &  $\min$ & med & avg & $\max$  \\ \midrule
 \texttt{small} & 34 & 115 & $222.52$ & 620 & 62 & 612 & $520.61$ & 930 \\
\texttt{medium} & 235 & 1,302 & $1,448.44$ & 4,941 & 1,017 & 3,032 & $3,343.44$ & 8,581 \\
\texttt{big} & 1,224 & 7,610 & $7,963.64$ & 16,264 & 10,445 & 21,000 & $19,519.00$ & 47,594 \\
\bottomrule
\end{tabular}
\label{t:ex-statistics}
\end{table}

\section{Effectiveness of the Heuristic Improvements}
\label{ch:exper:effectiveness}

As described in Chapter \ref{ch:heur}, the heuristic improvements are the relevant mechanism that really speeds up our algorithm to make it usable in practice. Therefore, we want to start by investigating the power of the several improvements that we have developed. First, we describe the general impression obtained by non-automated, manual testing which then led us to an automated setup showing the performance on a larger dataset.

To aid the reader, we provide a short recap of the used improvements:

\newcommand{\bsp}{\texttt{b-sp}\xspace}
\newcommand{\bcpl}{\texttt{b-cpl}\xspace}
\newcommand{\bfi}{\texttt{b-fi}\xspace}
\newcommand{\dms}{\texttt{d-ms}\xspace}
\newcommand{\dsp}{\texttt{d-sp}\xspace}
\newcommand{\cdist}{\texttt{c-dist}\xspace}
\newcommand{\cpl}{\texttt{c-pl}\xspace}

\vspace{-1em}

\begin{description}
	\item \textbf{Recognizing infeasible instances:}
	 \begin{description}
			\item[b-cpl] Whenever a checkpoint list contains more than $\ell/2+1$ checkpoints, at least two consecutive checkpoints need to be adjacent in the graph (Section \ref{h:b-cpl}).
			\item[b-sp] The sum of lengths of shortest paths between consecutive checkpoints cannot exceed $\ell$ (Section \ref{h:b-sp}).
		\end{description} 
		Whenever one of the conditions described above is violated, the instance is infeasible and we can return this fact. 
		\item \textbf{Breaking symmetries:}
	 \begin{description}
			\item[b-fi] Whenever a subtree of the search tree created by inserting $u$ as a checkpoint between $a$ and $b$ fails, $u$ can never be inserted between $a$ and $b$ in any of the sibling trees (Section \ref{h:b-fi}).
		\end{description} 
	\item \textbf{Improving the greedy approach: }\begin{description}
			\item[d-ms] After greedily fixing the $i$-th possible $s$-$t$ path, the minimum $s$-$t$ separator in the remaining graph cannot be smaller than $k-i$. If it is not and the remaining checkpoint lists only contain $s$ and $t$, then we can branch using Branching Rule 3 (Section \ref{h:d-ms}).
		\end{description}
	\item \textbf{Choosing the next child problem to expand:} \begin{description}
			\item[c-dist] When we branch and want to insert some vertex into a checkpoint list between vertices $u$ and $v$, we choose the vertices in ascending order of their distance to $u$ and $v$, i.e. we choose the vertex that is closest to both $u$ and $v$ first (Section \ref{h:c-dist}).
			\item[c-pl] When we branch in Branching Rule 2 and we have to choose a ``faulty'' subpath to branch over, we choose the subpath in descending order of their length, i.e. we choose the longest subpath first (Section \ref{h:c-pl}).
		\end{description}
\end{description}

\subsection{Methodical Setup}

\begin{figure}[t]
  \centering
  \includegraphics[width=.9\textwidth]{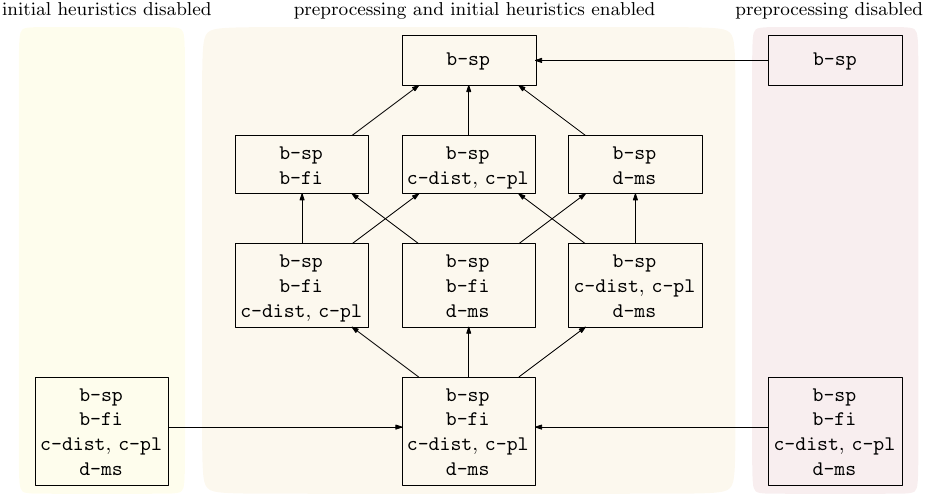}
  \caption{The different configurations tested in the experiments. The arrows indicate against which reference value the configuration is compared to.}
  \label{fig:res:heurtest-setup}
\end{figure}

To get an intuition for the effectiveness of the improvements, we started by exploring different configurations and how they perform on some of the test instances. After some manual experimenting, it quickly turned out that some of the improvements are necessary to allow the others to really show an effect. 

\subsubsection{Baseline configuration}

First of all, running the algorithm without any improvements showed to be not feasible on our dataset in a reasonable amount of time. Almost the full search tree would have to be expanded, which results in a very large runtime, exceeding our execution timeout of 10 minutes in a large majority of our test instances. Therefore, measuring the ``raw'' runtime of the search-tree algorithm had to be done in a different way.

To get reasonably fast running times, neither the improvements that choose subproblems intelligently nor those that decrease the branching factor help much if the depth of the search tree is huge. We need to have a reasonably small search tree to effectively employ those further techniques. This brings us to starting with some of the three \texttt{b-}improvements. Unfortunately, similar to running without any improvements, the first infeasible instance detection mechanism~\bcpl turned out to not really be useful, even though interesting from a theoretical perspective. Intuitively, this rule can only fire when a checkpoint list length of at least $\ell/2+1$ is reached, which still allows for a very large expansion of the search tree until it has an effect. In practice it therefore could not yield acceptable runtimes. On the other hand, the slightly more runtime-intensive, stricter rule \bsp provided a very good starting point with often acceptable runtimes, which is why we consider it as a baseline for all the remaining experiments.

\subsubsection{Improvement configurations to evaluate}

Building on this basis, there are three types of improvements that we wanted to test in multiple combinations: The \bfi rule for breaking symmetries by introducing forbidden intervals, the \cdist and \cpl heuristics for intelligently choosing subproblems and \dms for introducing an additional failure condition to the greedy runs.

We tried all of those separately together with \bsp to see how they alone affect the performance. Then, we tried out the three possibilities of combining two of them to see if they mutually influence their effect. Lastly, we ran all of the mentioned improvements together to see what effect this has.

Furthermore, whenever an instance got solved immediately by some initial pre-search-tree heuristic (Chapter \ref{ch:heur:prealg:minsep}), we ran it once without it to see how the search-tree algorithm would perform on those seemingly trivial instances. Lastly, we also wanted to test the effect of the preprocessing (Chapter \ref{ch:heur:preproc}) on the performance of the algorithm – therefore, we ran with disabled preprocessing the first configuration only consisting of \bsp, and the configuration of all improvements, to directly compare the effect. 

Figure \ref{fig:res:heurtest-setup} gives an overview of the different configurations used. As we experienced some bias towards the first run of any instance taking longer and the subsequent calls having a tendency of being faster, we randomized the execution order of the configurations to eliminate this bias as much as possible.

\subsubsection{Sample of the dataset}

To representatively show the performance gains of the different configurations, we need a good sample of instances to test them on. For this, we first solved a large number of instances consisting of randomly chosen $s$-$t$ pairs on all graphs of the dataset with all improvements enabled, which seemed to be the fastest in preliminary testing. Of those we then sampled for each combination of $k \in [2,7]$ and $\ell \in [5,9]$ one positive and one negative instance having a search tree size of $(0,100], (100,1000]$ and $(1000,10000]$ nodes to also be able to judge the effect of the improvements on ``easier'' instances compared to those which are even quite hard with all improvements enabled. Additionally, we sampled $10$ instances for each graph which were trivially solved by some of the pre-search-tree heuristics. 

In one exception, we did not use this sample – as mentioned before, the ``bare'' algorithm without any improvement enabled did not terminate on our instances in most of the cases, which made it hard to find reference values to compare our baseline improvement \bsp to. Therefore, in this exception we randomly generated small graphs on which the bare algorithm does terminate – the approach is described in the result section.

\pagebreak

\subsection{Results}

The results can be viewed in multiple ways. Firstly, we consider the relative changes of search tree size or runtime for each specific instance compared to some given baseline. Those relative changes correspond to dividing the value (of nodes/of runtime) required with some improvement by the corresponding value required by the baseline. Resulting values in the range between 0 and 1 therefore indicate an improvement caused by the improvement, values above 1 a worsening. Furthermore, one can also consider the average runtime per search tree node for different improvements, which should give a hint on how computationally expensive the different improvements are.

In the following sections, we want to discuss the results for the different improvement configurations.

\subsubsection{Base configuration: \bsp}

\begin{figure}[t]
  \centering
  \includegraphics[width=\textwidth]{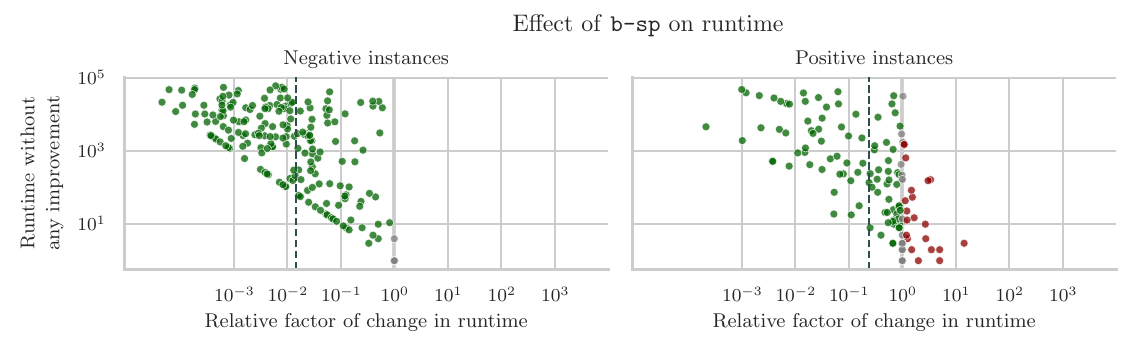}
  \caption{Scatter plot showing on the x-axis the relative changes of the runtime using one \bsp compared to using no improvement, and on the y-axis the initial runtime without using any improvement. The dashed line marks the geometric mean of the factor of change.}
  \label{fig:res:bsp-time-rel}
 \end{figure}

As mentioned in the section about the methodics, using the ``bare'' algorithm without improvements did not yield usable results for our regular dataset – in the large majority of the instances, our timeout of 10 minutes was exceeded. Therefore, this way we also were not able to judge the effect of using \bsp, as we had no baseline. This showed that this simple improvement, where we compute shortest paths between checkpoints in the full graph to check whether it would even make sense to try to satisfy this checkpoint list, has a significant effect. However, on our given dataset sample we had no way to quantify that effect. 

For this reason, to show the performance of the base configuration \bsp, different from our usual methodics, we randomly generated graphs using the Erdős–Rényi–Gilbert model \cite{Gilbert1959} generating a graph $G(n,p)$ with $n$ vertices and each edge occuring with a probability of $p$. We chose values of $n \in [20,40,60]$, $p \in [0.05,0.1,0.15,0.2,0.3]$ and for our instances we always tried to find $k \in [2,4,6]$ paths of length at most $\ell \in [5,7,9]$ between the same two vertices. As those instances are smaller, the algorithm terminates much sooner, allowing us to measure ``raw'' runtime without improvements. Nevertheless, it is very hard to find combinations of those values which neither produce mostly timeouts, nor results in only trivial instances which are recognized by one of our initial heuristics. With some of the combinations of parameters however, we were successful.

Figure \ref{fig:res:bsp-time-rel} shows the effect of using \bsp on our randomly generated nontrivial instances. The logarithmic x-axis represents the factor of change in runtime that the specific improvement causes. Whenever the runtime decreases, the points are colored in green, when the runtime increases, they are colored in red. If it (roughly) stays the same, the dot is colored gray. The logarithmic y-axis on the other hand depicts the initial value of the runtime required by the reference configuration, in our case the ``bare'' algorithm. Therefore, a green dot in the upper left corner corresponds to an instance, where without improvements the algorithm required a large search tree to decide the instance, but now with the improvement enabled the search tree shrinks to a factor of less than $10^{-3}$. Furthermore, the instances are separated in negative and positive instances to better differentiate the effects of the improvement and the geometric mean of the relative factor of change is marked by a dashed line.

We can see, that \bsp indeed has a highly significant effect forming a nearly linear function on a log-log scale, corresponding to a double exponential improvement. The effect is similar for negative as well as positive instances, even though for negative instances there is not a single worse runtime, while for positive instances, especially in the instance with smaller initial runtime, there are a few examples of runtimes getting longer.

This can be explained by the runtime overhead introduced by computing shortest paths in the graph. For instances, that were already quite fast without \bsp, this means that the overhead of \bsp is not able to save more runtime in reducing search tree nodes than it additionally requires for the shortest paths computations. In general, however, we can see that the effect of \bsp is on average between one or two orders of magnitude. So for the coming results, we have to always keep in mind that a large part in improving the runtime is due to \bsp, even though it only acts as a reference value in the rest of this chapter. 

\subsubsection{Single improvements: \bfi\ / \cdist+ \cpl\ / \dms}
\begin{figure}
  \centering
  \includegraphics[width=\textwidth]{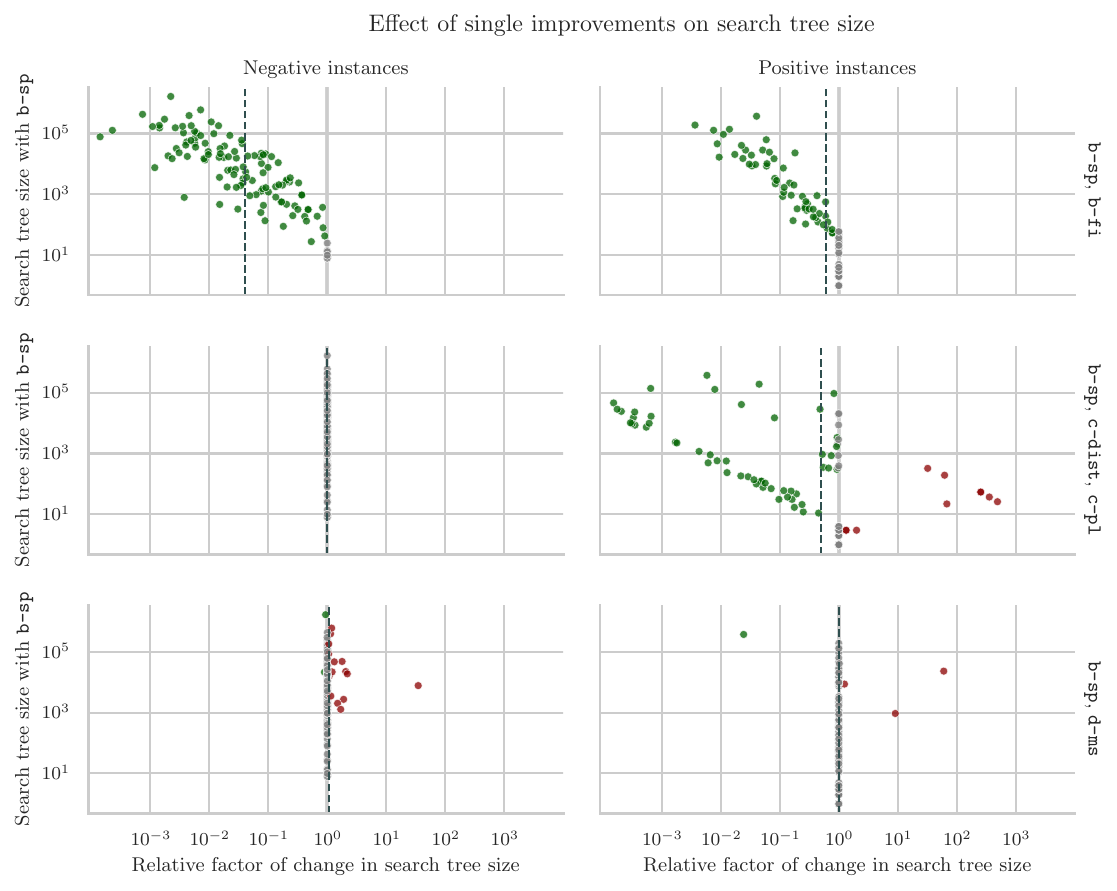}
  \caption{Scatter plot structured the same way as \ref{fig:res:bsp-time-rel}, but showing on the x-axis the relative changes of the search tree size using one of the improvements \bfi, \cdist/\cpl or \dms in addition to \bsp, and on the y-axis the initial search tree size when only using \bsp. The dashed line marks the geometric mean of the factor of change.}
  \label{fig:res:single-nodes-rel}

\end{figure}

We now come back to our dataset and the methodics explained in the previous section, and start by looking at how the improvements \bfi, the \texttt{c}-heuristics and \dms improve the performance when used together with our baseline \bsp. Figure \ref{fig:res:single-nodes-rel} shows the effects of the application of the different improvements on a scatter plot compared to only using the baseline improvement \bsp. It is plotted in the same way as Figure \ref{fig:res:bsp-time-rel}, but instead of considering only the runtime, we look a bit more into detail and start by comparing the search tree size, i.e. the number of nodes.
The logarithmic x-axis still represents the factor of change that the specific improvement causes, but now for the search tree size and the logarithmic y-axis on the other hand now depicts the initial value of the search tree size required by the reference configuration.

Looking at Figure \ref{fig:res:single-nodes-rel}, we can see two improvements performing very well: The forbidden intervals introduced by \bfi have a consistent positive effect seeming to become stronger the larger the size of the initial search tree was. It has good effect on positive instances, but seems to have an even better effect on negative instances. The subproblem selection heuristics \cdist and \cpl have no effect on negative instances, while causing a nearly consistent improvement on positive instances which exponentially increases with the initial search tree size. The greedy localization improvement \dms on the other hand has very little effect, in some cases even growing the search tree size.
	
\begin{figure}
		\centering
         \centering
         \includegraphics[width=\textwidth]{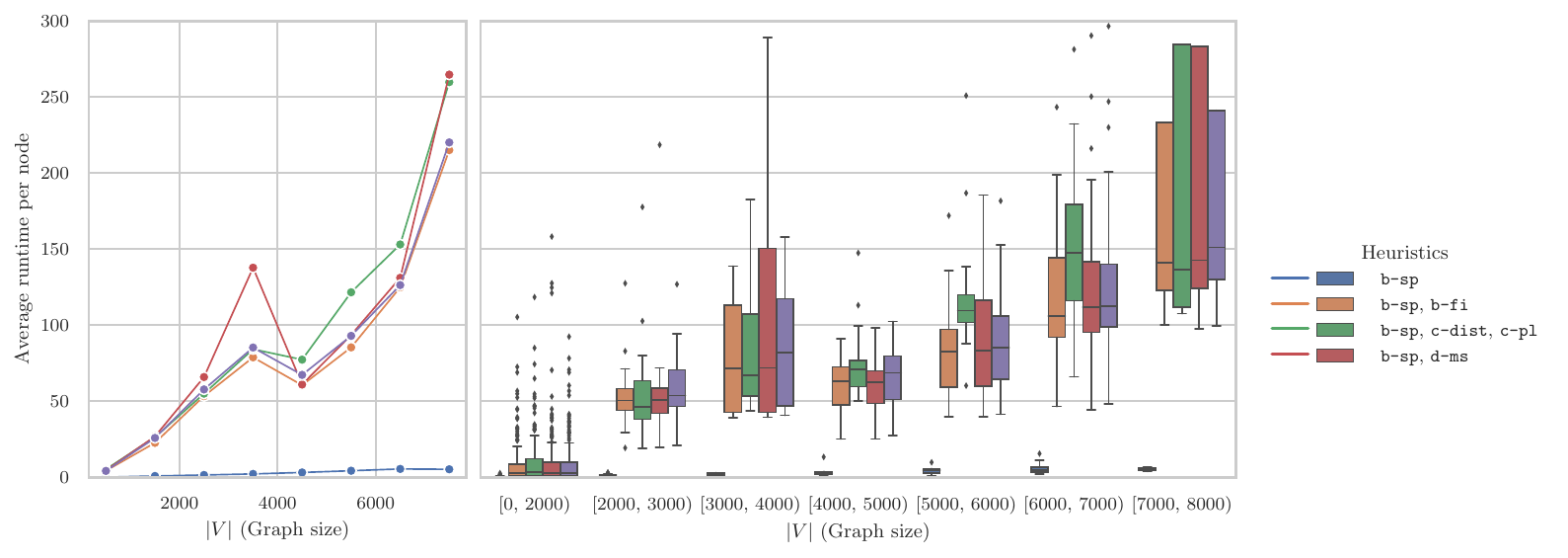}
  		   		 \vspace{-1.6em}
     \caption{Plots showing the average runtime per search node for the different improvements, grouped by bins of different graph sizes. On the left, the mean values are shown in a lineplot, on the right, median and quartiles in a boxplot.}
     \label{fig:res:single-avg-combined}

\end{figure}

The interesting question however is how this affects the runtimes, as search tree nodes are not everything – if we halve the number of search tree nodes, but the computations need double the time per search tree node, we gain nothing. To get a clear picture of how computationally intensive each optimization is, we view the average per-node runtimes. As those highly depend on the polynomial-time algorithms employed within the search tree nodes, the runtime depends on the size of the graph. Figure \ref{fig:res:single-avg-combined} shows the statistics for average runtimes against binned graph sizes. There we can see, that the \texttt{c}-heuristics and \bfi seem to have the highest computational overhead, compared to those of \dms and the base configuration \bsp seeming to be reasonably small.

\begin{figure}
  \centering
  \includegraphics[width=\textwidth]{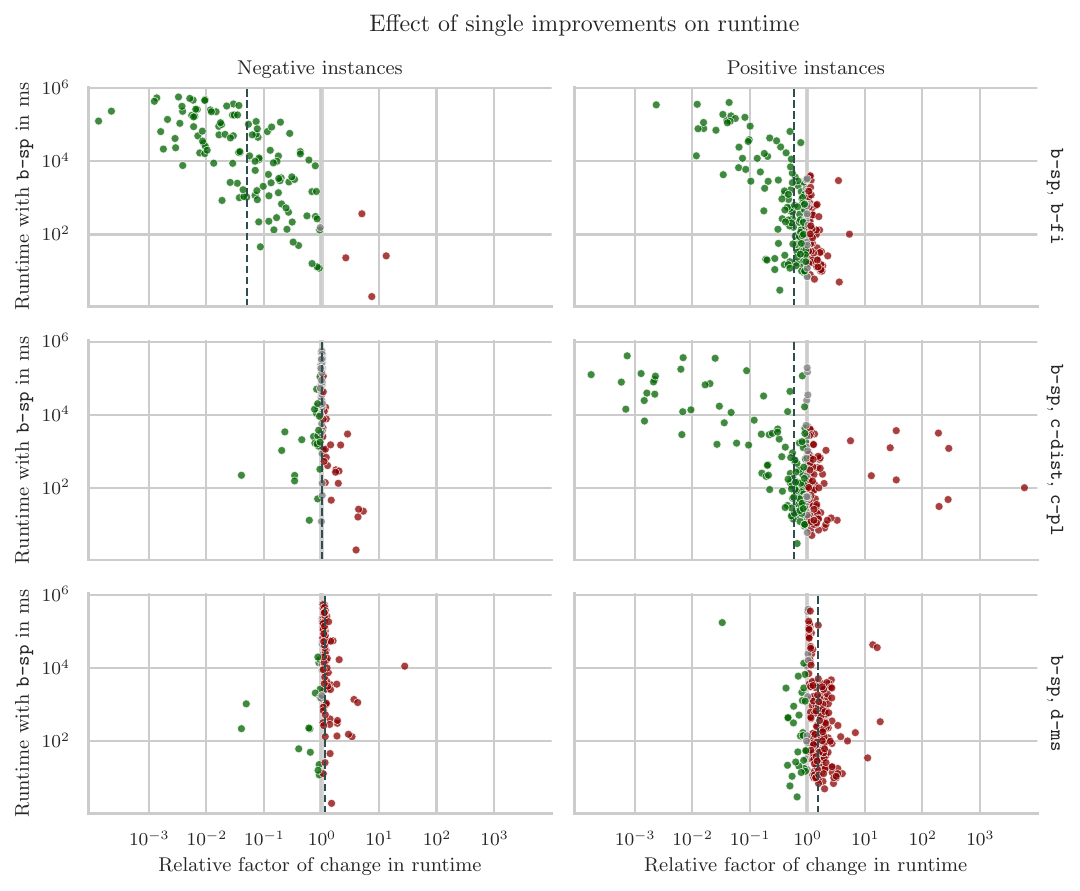}
  \caption{Scatter plot structured the same way as Figure \ref{fig:res:single-nodes-rel} showing the relative changes of the runtimes (in miliseconds) when using the different improvements.}
  \label{fig:res:single-time-rel}
\end{figure}

Putting those together, Figure \ref{fig:res:single-time-rel} plots the relative changes of the runtimes the same way as done before for the number of nodes in the search tree. The logarithmic x-axis still shows the factor of change, while the logarithmic y-axis now shows the initial runtime in milliseconds without using the corresponding improvement. For \bfi, we can see that the positive effect on the nodes translates to a better runtime, even though it has the most computational overhead compared to the other configurations. This overhead shows in the instances having a lower initial value, where worse runtimes occur. For the \texttt{c}-heuristics, we can see that the effect on negative instances is very limited with no tendency, while the positive instances experience an overall improvement with only few instances getting worse. The greedy improvement \dms has limited effect leaning more towards making the runtime worse, but not by large orders of magnitude.

Those observations correspond to what we expected: The forbidden interval rule \bfi is a very strong mechanism for reducing the search tree size. Even though it seems to be computationally more expensive, which most likely comes from the overhead of maintaining and checking the forbidden interval lists, it still achieves large improvements, especially in negative instances having a high initial runtime.

The \texttt{c}-heuristics decide over the next subproblem to expand. Intuitively, for negative instances the whole search tree has to be traversed and therefore the order of subproblems should effectively not make any difference. This is confirmed in our data. However, we can also see that this is still just a heuristic: In a lot of cases, the order used without this heuristic turned out to be ``lucky'' and gave the successful result much faster. Still, the positive effects dominate.

\begin{figure}
\vspace{-.5em}
	\centering
    \begin{subfigure}[b]{.85\textwidth}
         \centering
  \includegraphics[width=\textwidth]{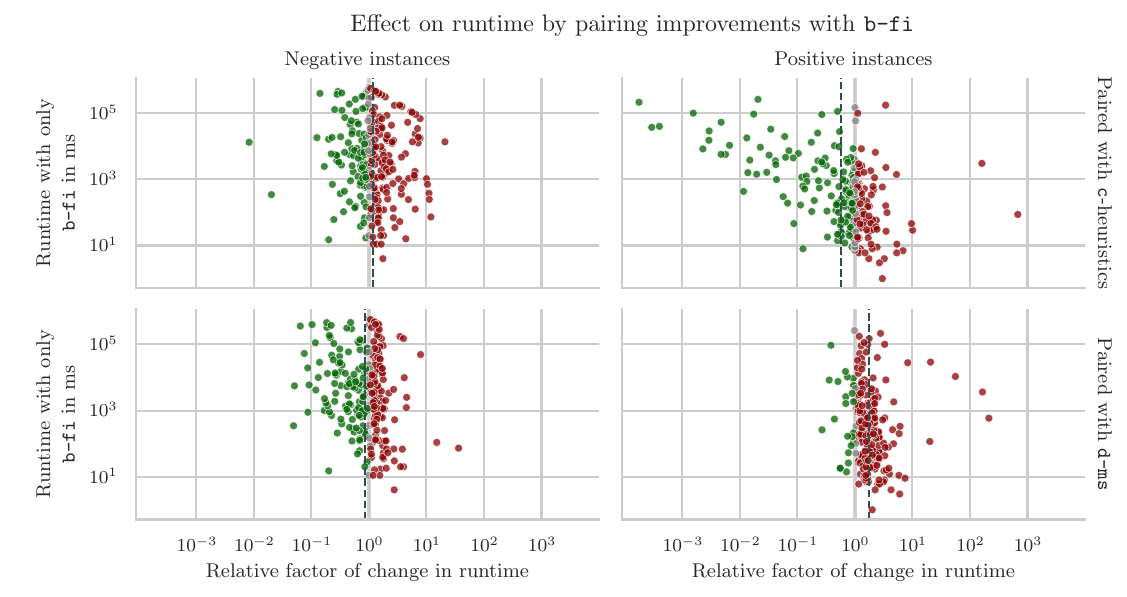}  
         \vspace{-1.8em}

  \caption{\bfi together with \texttt{c}-heuristics (top) and \dms (bottom)}
             \label{fig:res:double-time-a-rel}
       \vspace{.7em}

     \end{subfigure}
	\begin{subfigure}[b]{.85\textwidth}
         \centering
  \includegraphics[width=\textwidth]{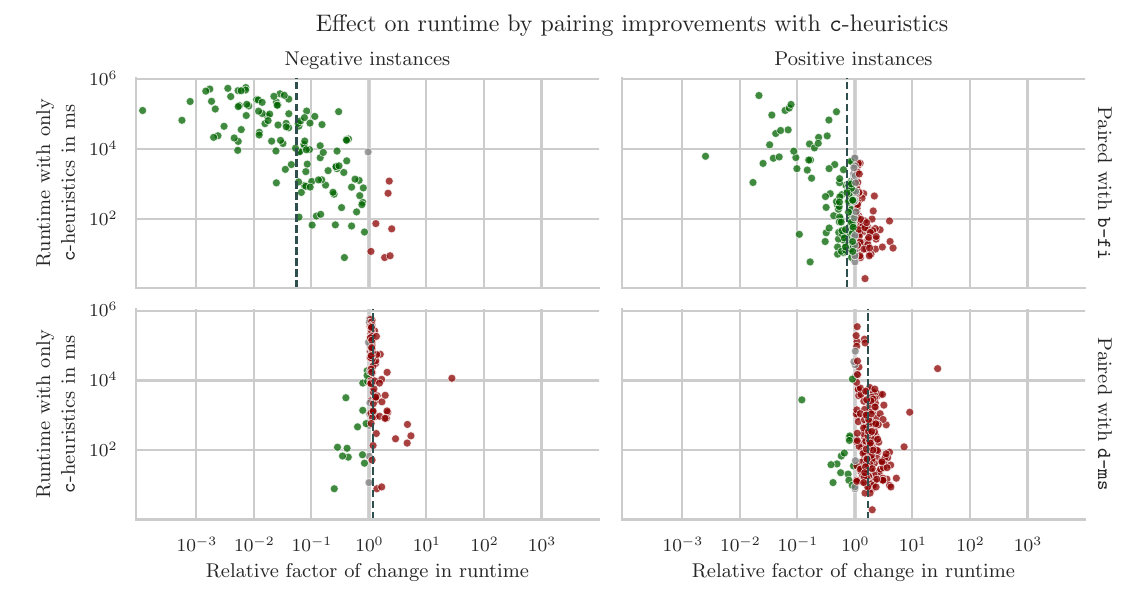}  
           \vspace{-1.8em}

  \caption{\texttt{c}-heuristics together with \bfi (top) and \dms (bottom)}
             \label{fig:res:double-time-b-rel}
                       \vspace{.7em}
     \end{subfigure}

    \begin{subfigure}[b]{.85\textwidth}
         \centering
  \includegraphics[width=\textwidth]{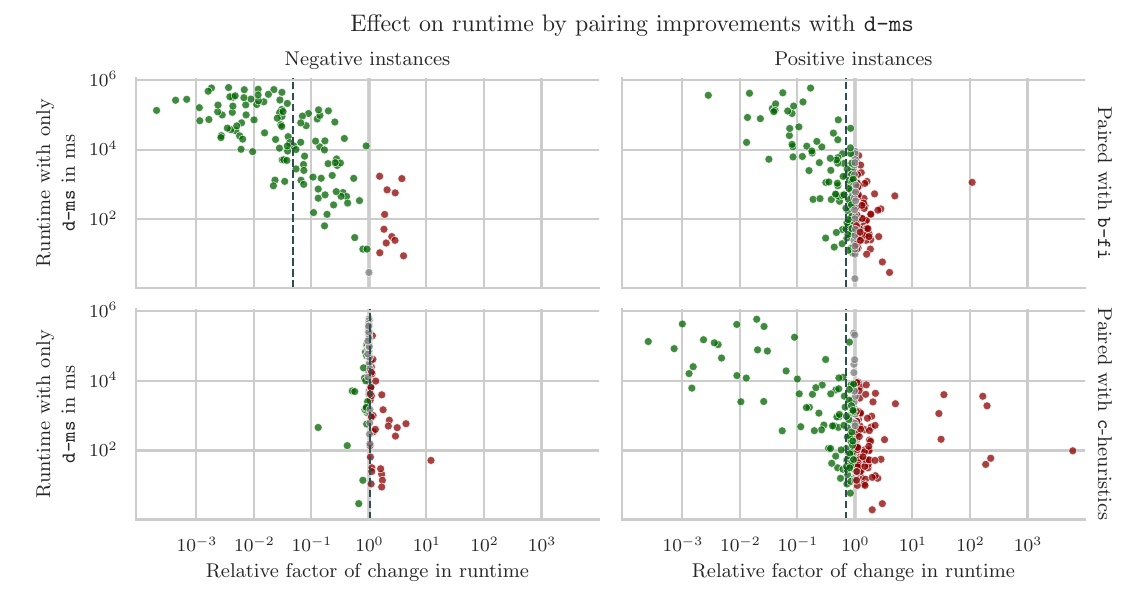}  
           \vspace{-1.8em}
  \caption{\dms together with  \bfi (top) and \texttt{c}-heuristics (bottom) }
             \label{fig:res:double-time-c-rel}

     \end{subfigure}
     	
           \caption{Scatter plots structured the same way as Figure \ref{fig:res:single-nodes-rel} showing for each improvement the relative factor of change to the runtime caused by additionally enabling one of the two remaining.}
           \label{fig:res:double-time-rel}
\end{figure}

For the greedy improvement \dms, we had mixed expectations. As it involves the use of a new branching rule with a higher worst-case branching factor, but at the same time allows for earlier recognition of infeasible sets of paths, this tradeoff was the most interesting part to investigate experimentally. We observed that in over 60 percent of the instances \dms fired at least 1000 times. However, in over 90 percent the runtime did not get worse than double. Therefore, it seems like it is able to keep the balance in a way that using it seems to have neither positive nor crucially negative effects on the runtime.

\subsubsection{Paired improvements: combining the approaches}

Based on the promising results above we want to combine the improvements to see if this produces even better results. To investigate the influence the improvements have on each other when used together, we tested all three pairwise combinations of the single improvements evaluated in the previous section.

Figure \ref{fig:res:double-time-rel} shows the relative factor of change in runtime for each improvement when additionally enabling one of the two other. For \bfi together with the \texttt{c}-heuristics (Figure \ref{fig:res:double-time-a-rel}), we can see that the effect on the positive instances is even stronger, while for the negative instances, there are almost the same number of instances experiencing better and worse runtimes, keeping the balance. The improvement \dms, contrary to when being used alone, shows to have positive effect on negative instances, while keeping the trend of prolonging the runtimes for positive instances.

Using the \texttt{c}-heuristics as a baseline (Figure \ref{fig:res:double-time-b-rel}), we can see that adding \bfi has an almost exclusively positive effect on negative instances and in many cases also for positive instances, while the general trend for positive instances seems to be that the runtimes stay roughly the same. Adding \dms however does not seem to make a huge difference with the tendency of making the runtimes worse.

Plotting the other improvements against \dms (Figure \ref{fig:res:double-time-c-rel}) does not give us much information, as we could already not see any noteworthy effect when using it alone. Therefore, adding in both of the other improvements mainly gives us the savings of \bfi and the \texttt{c}-heuristics which we already saw when using them alone (Figure \ref{fig:res:single-time-rel}).

The most interesting observation that can be made by looking at the plots in Figure~\ref{fig:res:double-time-rel} is the fact that, different to the single improvement tests, now \dms seems to show a good effect on negative instances when used together with \bfi. Interestingly, the \texttt{c}-heuristics, which intuitively only should reduce the time necessary to find positive instances, now together with \bfi also have an effect on some of the negative instances. 

An explanation for both of those effects can be found by reminding ourselves again about the forbidden interval rule. There, we keep track of already-used unsuccessful checkpoint insertion configurations (which we call forbidden intervals), which then can be used in sibling search tree nodes and their descendant nodes to rule out infeasible instances earlier on. 

For \dms, we use a modified branching rule which creates up to $k$ times more child instances. Our (speculative) explanation for \dms performing so well together with~\bfi is the higher number of child instances in some cases allows more of the forbidden intervals to accumulate, having a stronger effect in the later sibling nodes, therefore limiting the search tree size more effectively.
\enlargethispage{2\baselineskip}
Regarding the \texttt{c}-heuristics, the choice of the next subproblem also determines which forbidden intervals get created. It seems that our heuristics here are not only useful for determining search paths leading to positive instances, but also for negative instances finding good paths for effective forbidden intervals.

In general, we can see that none of the combinations cause a significant increase in runtime and therefore it is definitely advisable to use multiple improvements together.

\subsubsection{All improvements: putting everything together}

\begin{figure}
  \centering
  \includegraphics[width=.85\textwidth]{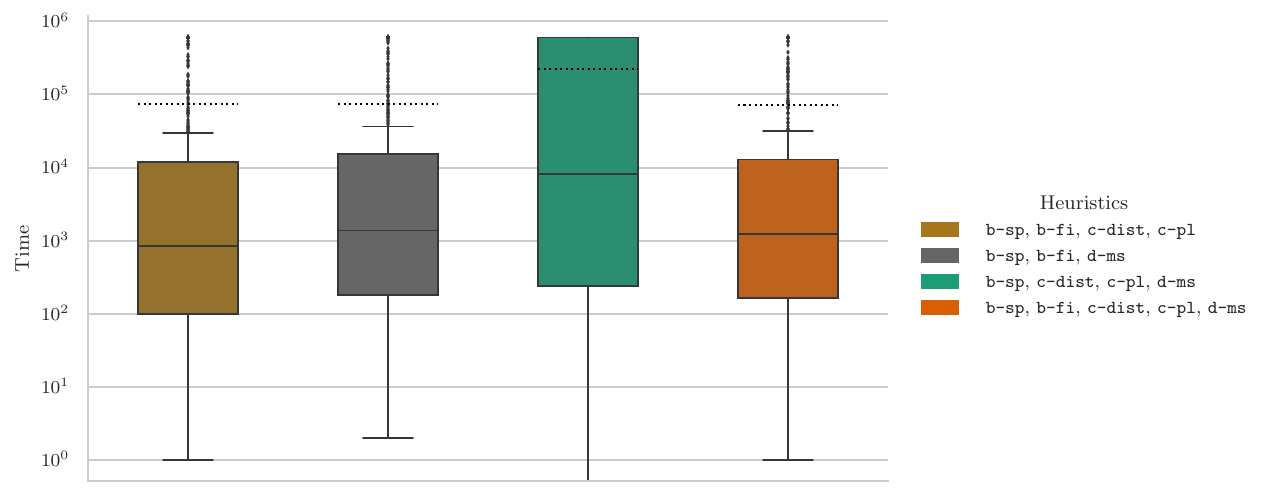}
  \caption{Boxplot showing the distribution of the runtimes for the different improvement configurations, additionally the arithmetic mean is marked by a dashed line.}
  \label{fig:res:full-time-boxplot}
  
  \includegraphics[width=\textwidth]{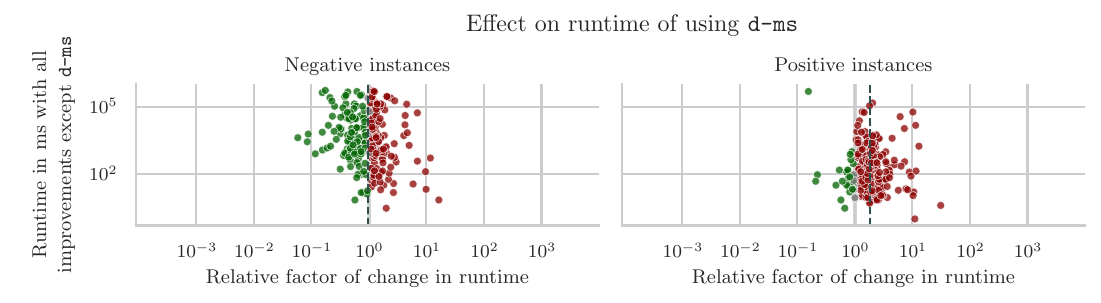}
  \caption{Scatter plot structured the same way as Figure \ref{fig:res:single-nodes-rel} showing the relative changes of the runtimes when using \dms together with all other improvements against not using it.}
  \label{fig:res:full-time-c-rel}
\end{figure}

Finally, we consider the combination of all three improvements together and compare it to the pairwise combinations from the previous section.

Figure \ref{fig:res:full-time-boxplot} shows a boxplot for the runtimes of the different configurations. In terms of the median and the quartiles, the full configuration seems to be similarly as good as the one with \bfi and the \texttt{c}-heuristics. Regarding the mean, the full configuration seems to have a slight advantage (\SI{71,688}{ms} vs.\ \SI{73,041}{ms}). When comparing those two configurations directly (Figure \ref{fig:res:full-time-c-rel}), we can see especially for harder, negative instances there is a tendency to longer running times, while the increases happen more on the easier instances. 

This result leads us to the conclusion that when aiming for the shortest expected runtime, it makes sense to use the full configuration combining all improvements, even though there exist many easier instances where the individual runtime gets worse by adding in \dms.

\subsection{Preprocessing}

\begin{figure}
	  \centering
  \includegraphics[width=\textwidth]{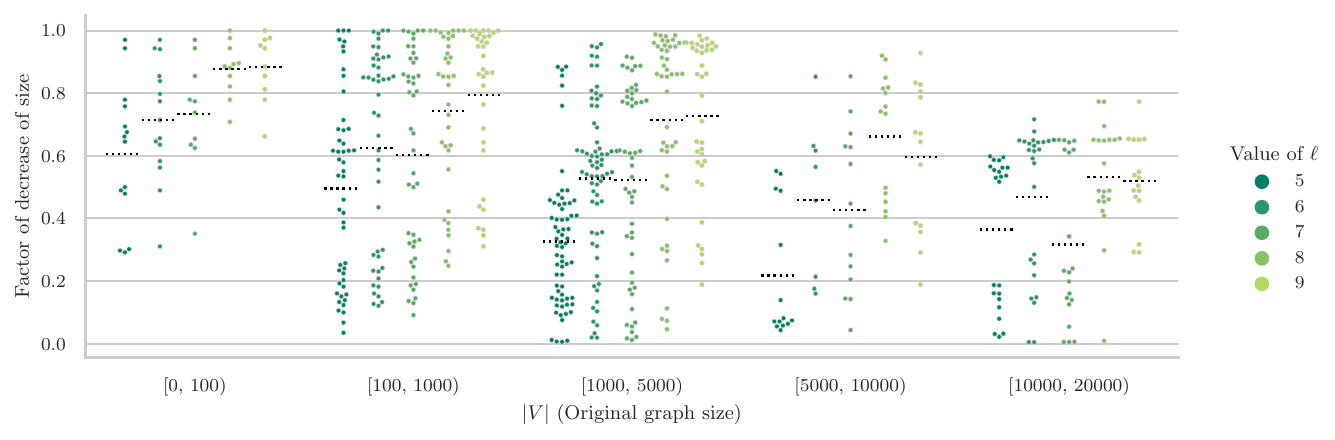}
  \caption{Swarmplot showing the decreases in graph sizes for different size classes and different values of $\ell$, the mean value is shown as a dashed line.}
  \label{fig:res:heur-preproc-size}
\end{figure}

Another different aspect of our heuristic improvements is the preprocessing of the graph described in Chapter \ref{ch:heur:preproc}. To measure its effect, we consider two aspects: How effectively reduced do the graphs get for different parameters and how does this affect the runtime?

\begin{figure}

  \begin{subfigure}[b]{\textwidth}
         \centering
  \includegraphics[width=\textwidth]{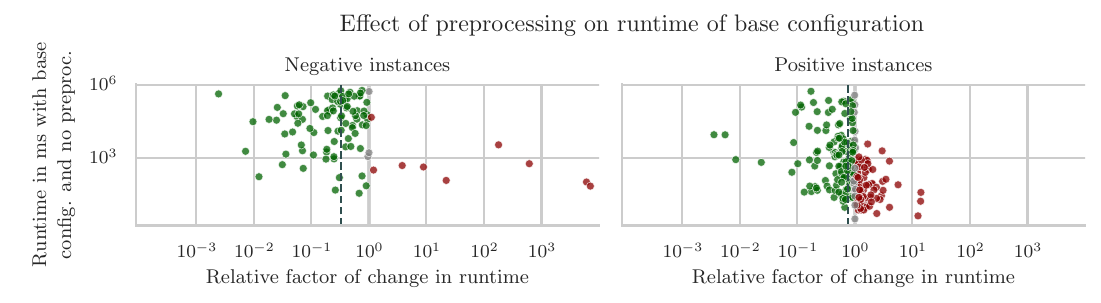}  

  \caption{Runtimes for only using \bsp comparing with and without preprocessing}
             \label{fig:res:heur-preproc-time-1}
     \end{subfigure}

    \begin{subfigure}[b]{\textwidth}
         \centering
  \includegraphics[width=\textwidth]{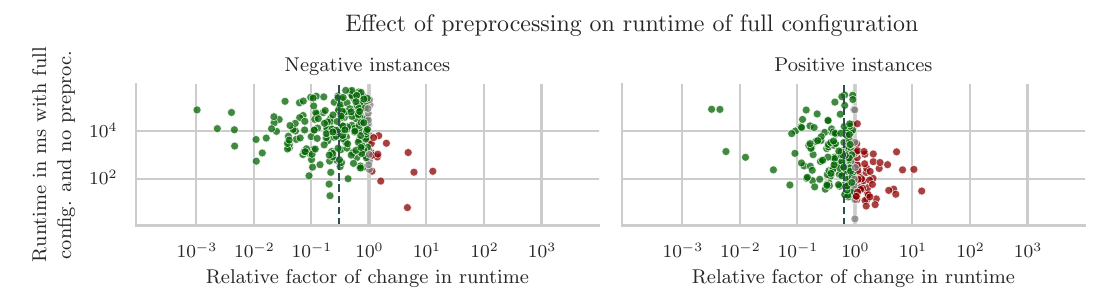}  
           \vspace{-1.5em}
  \caption{Runtimes for all improvements enabled comparing with and without preprocessing}
             \label{fig:res:heur-preproc-time-2}

     \end{subfigure}
\vspace{-1em}
     \caption{Scatter plots structured the same way as Figure \ref{fig:res:single-nodes-rel} showing the effect of the preprocessing on the runtime.}
     \label{fig:res:heur-preproc-time}
\end{figure}

In Figure \ref{fig:res:heur-preproc-size}, we look at the factor of decrease of the number of vertices for different graph sizes and different $\ell$. The value on the y-axis tells us the remaining percentage of vertices in the reduced graph. We can see that for small $\ell$-values, the reduction is stronger, and in general the preprocessing shows more effect on the larger graphs. The runtime increases/decreases are displayed in Figure \ref{fig:res:heur-preproc-time} – we first compare running only \bsp (Figure \ref{fig:res:heur-preproc-time-1}), then all improvements (Figure \ref{fig:res:heur-preproc-time-2}), each with and without preprocessing. There, we observe that there is a significant positive effect. For positive instances, the preprocessing almost exclusively causes improvements, while for negative instances there are some worsenings for lower initial runtimes. Contrary to the runtime, the size of the search tree does not change at all by the preprocessing (not shown in any figure).

In general, the factor of reduction grows with the size of the graphs – this is what we expected, as for large graphs, there is simply more potential of removing vertices. Similarly, the reason for smaller values of $\ell$ resulting in smaller graphs is clear: Our graphs always shrink down to some $O(\ell)$-neighborhoods of $s$ and $t$, and therefore a smaller $\ell$ allows us to remove more vertices from the graph. The effect on the search tree nodes and the runtime is interesting: As already briefly mentioned, the number of search tree nodes stays exactly the same when using the preprocessing – and knowing this, the gains in runtime appear quite strong. While for the other improvements, the effect was mainly based on the reduction of search tree nodes, here all improvements are solely caused by all polynomial-time algorithms employed within the algorithm being much faster because of the reduced graph size. 

\subsection{Initial Heuristics}

Lastly, a significant part of our improvements, even though technically not part of the algorithm itself, are our initial so-called pre-search-tree heuristics for detecting trivial instances. We recall that they highly depend on two algorithms: One is Suurballes algorithm for finding disjoint $s$-$t$ paths of minimum total length, and the other is an algorithm for determining minimum $s$-$t$ vertex separators in the graph. The former is used to detect trivial positive and negative instances: When we have a solution of $k$ paths of minimum total length, where the longest path is shorter than $\ell$, then this solution also applies to \SPp. On the other hand, when the minimum total length exceeds $k \cdot \ell$, then we know a solution to \SPP also cannot exist. The latter is even more straightforward: If the minimum $s$-$t$ separator in the graph is smaller than $k$, then no solution can exist.

\begin{figure}[t]
  \centering
  \makebox[\textwidth][c]{\includegraphics[width=1\textwidth]{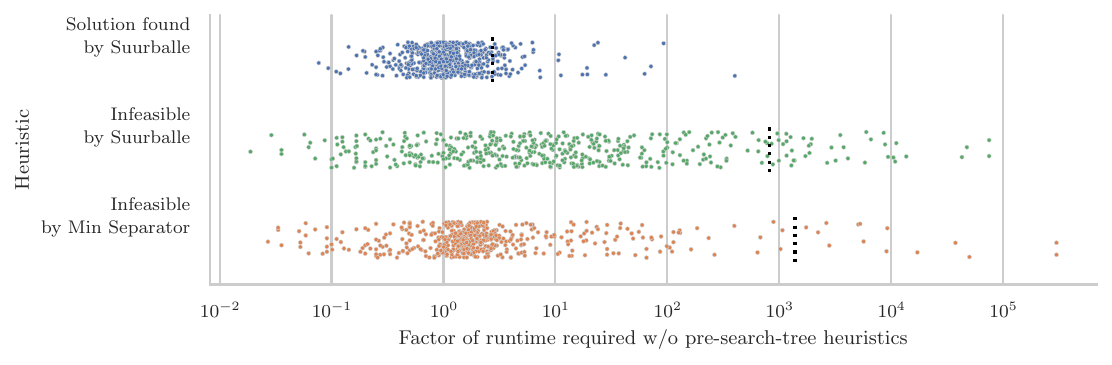}}
  \caption{Strip plot showing the runtimes for trivial instances with the pre-search-tree heuristics disabled, the mean value is marked by a dashed line.}
  \label{fig:res:preheur}
\end{figure}

Our experiments have shown that those initial heuristics fire with a very high frequency in practical use. Therefore, we want to analyze how much runtime they save compared to running instances without those heuristics enabled. Figure \ref{fig:res:preheur} shows for each of the mechanisms in place how long our algorithm would take if we disabled it. The positive instances do not seem to highly depend on Suurballes algorithm, as on average the runtime without using it is only three times as high. On the other hand, the negative instances recognized by Suurballe and the min separator heuristic on average take about 1000 times as long to decide.

Of course, there are many instances where even without those heuristics the algorithm terminates very fast. As we have high fluctuations in those short runtimes, we have quite a few samples where the runtimes without those heuristics seem to even be faster. However, one should not misinterpret this: Contrary to the previous sections, the initial runtimes here lie in the spectrum of at most \SI{1000}{ms}, as the initial heuristics only involve fast polynomial-time algorithms. Therefore, any ``decrease'' by a factor of 0.1 is difficult to distinguish from ``noise'' coming from the machine we took the experiments on, compared to an average increase of 1000, resulting in some instances reaching our timeout of 10 minutes, thus being highly significant. 

Those results give us a hint that our mechanisms in place for finding positive instances seem to be quite good even without employing Suurballes algorithm, as the runtimes do not increase significantly. On the other hand, the bad performance of our search tree for negative instances shows that there might be much more potential in improving the runtime of negative instances.

\section{Performance of the Algorithm}
\label{ch:exper:performance}

Having determined a good configuration of improvements, we now want to heavily test the resulting algorithm on the whole dataset for all possible combinations of $k$ and $\ell$. We first explain our methodical approach, then describe results regarding the runtime and finally present some statistics on the frequency of use of the several branching rules and improvements.

\subsection{Methodical Setup}
 
Motivated by Section \ref{ch:exper:effectiveness}, we use all of the improvements which we tested, namely \bsp, \bfi, \cdist, \cpl and \dms. To cover as many different graphs as possible, we ran our experiments on every graph from the dataset of Nadara et al. \cite{NadaraEtAl2018}. The instances were created by randomly sampling pairs of vertices and checking if they have a distance of at most 10. Then, for each such vertex pair, we ran the algorithm for all values of $k \in [2,7]$ and $\ell \in [5,10]$. We did this for each graph until we sampled 100 non-trivial instances, meaning that for at least one $k,\ell$ combination the number of search tree nodes was larger than 1, indicating that neither the initial heuristics nor the initial greedy approach resulted in a solution. For the runtime we set a timeout of 10 minutes, after which the algorithm aborts its execution. 

\subsection{Results}

\begin{table}[t]
\caption{Result Statistics}
\begin{tabular}{lrrrrrr}
\toprule
Group           & \multicolumn{2}{c}{\texttt{small}}     & \multicolumn{2}{c}{\texttt{medium}}    & \multicolumn{2}{c}{\texttt{big}}     \\
\cmidrule(r){1-1} \cmidrule(lr){2-3} \cmidrule(lr){4-5} \cmidrule(lr){6-7}
Total Runs         & \multicolumn{2}{l}{8,308,548} & \multicolumn{2}{l}{5,557,032} & \multicolumn{2}{l}{\phantom{1,}994,176} \\  \midrule

Positive Instances & 5,829,609      & 70.16\%     & 926,586      & 16.67\%     & 93,155     & 9.37\%    \\ \cmidrule(r){1-1} \cmidrule(lr){2-3} \cmidrule(lr){4-5} \cmidrule(lr){6-7}
\quad Suurballe          & 5,507,867      & 66.29\%     & 884,471      & 15.91\%     & 86,107     & 8.66\%    \\ 
\quad Search Tree        & 321,742        & 3.87\%      & 42,115         & 0.76\%      & 7,048       & 0.71\%     \\ \midrule
Negative Instances & 2,478,745      & 29.83\%     & 4,630,319        & 83.32\%     & 900,655     & 90.59\%    \\ \cmidrule(r){1-1} \cmidrule(lr){2-3} \cmidrule(lr){4-5} \cmidrule(lr){6-7}

\quad Min Separator      & 2,475,300        & 29.79\%     & 4,624,170        & 83.21\%     & 894,702     & 89.99\%    \\
\quad Suurballe          & 1,434           & 0.02\%      & 2,900          & 0.06\%      & 3,756         & 0.37\%     \\
\quad Search Tree        & 2,011           & 0.02\%      & 3,249          & 0.05\%      & 2,197         & 0.22\%     \\ \midrule
Timeout            & 194           & <0.01\%      & 127            & <0.01\%      & 366           & 0.04\%    \\
\bottomrule
\end{tabular}
\label{t:res:perf}
\end{table}

\begin{figure}[th!]
	\centering
    \begin{subfigure}[b]{\textwidth}
         \centering
  \includegraphics[width=1.2\textwidth, center]{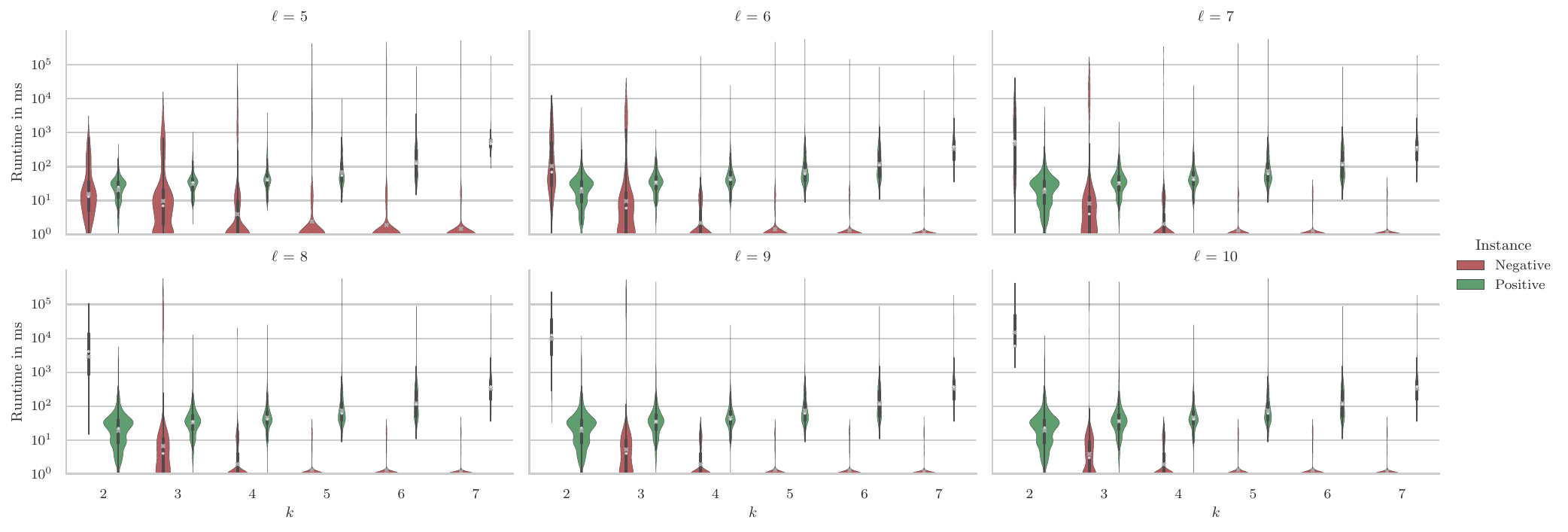}
         \vspace{-1.5em}

  \caption{Graphs from group \texttt{small}}
             \label{fig:res:perf-small}
       \vspace{1em}

     \end{subfigure}
	\begin{subfigure}[b]{\textwidth}
         \centering
  \includegraphics[width=1.2\textwidth, center]{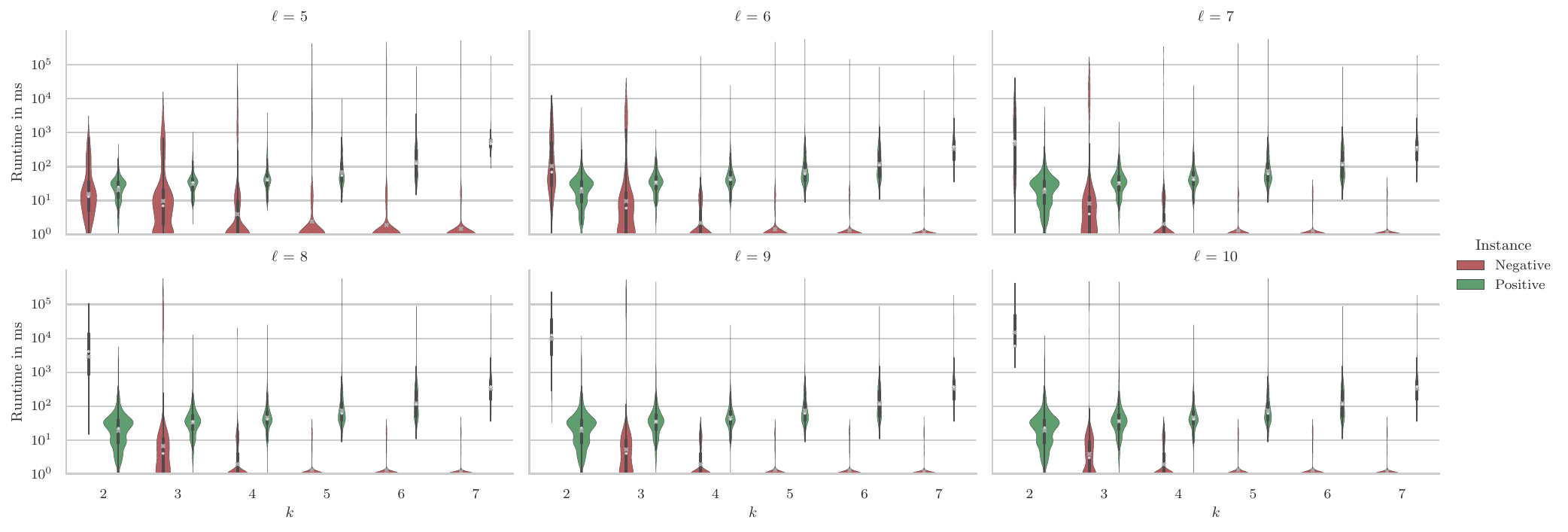}
           \vspace{-1.5em}

  \caption{Graphs from group \texttt{medium}}
             \label{fig:res:perf-medium}
     \end{subfigure}
          \vspace{.5em}

    \begin{subfigure}[b]{\textwidth}
         \centering
 \includegraphics[width=1.2\textwidth, center]{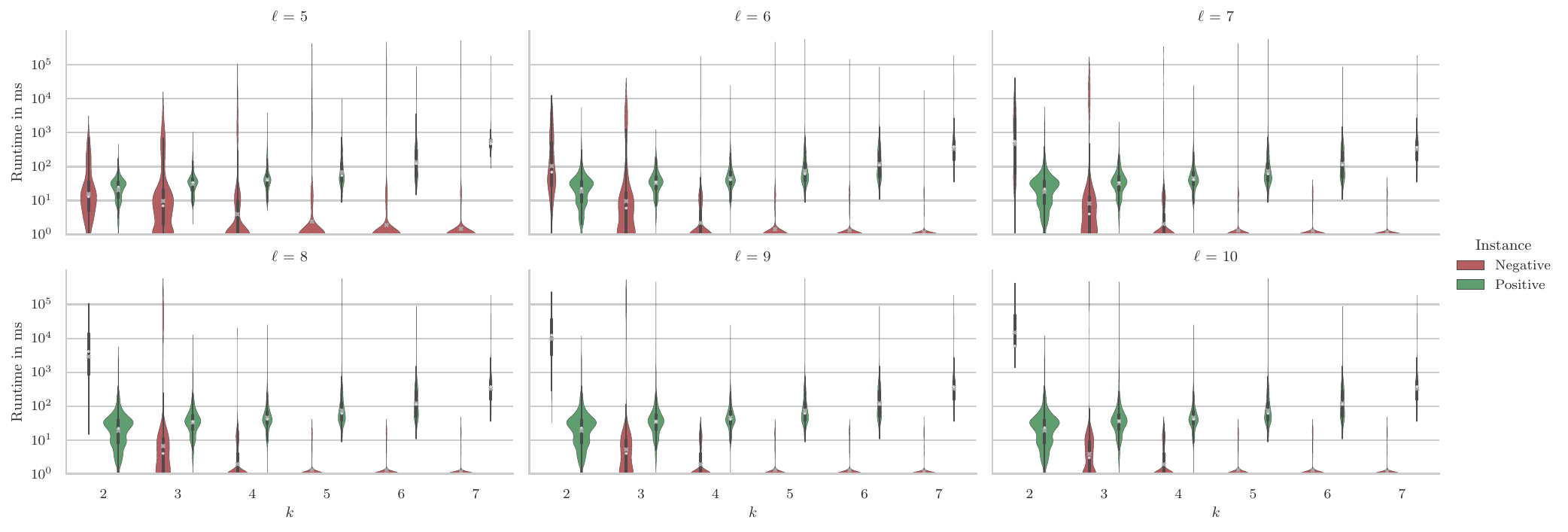} 
           \vspace{-1.5em}
  \caption{Graphs from group \texttt{big}}
             \label{fig:res:perf-big}

     \end{subfigure}
     	
           \caption{Violin plots showing the distribution of runtimes (in miliseconds) for each of the three dataset categories and all values of $k$ and $\ell$. Within each violin, a small boxplot is drawn and the arithmetic mean is marked by a cross.}
           \label{fig:res:perf}
\end{figure}

\subsubsection{Runtimes for different graph sizes and different $k$ and $\ell$}
Table \ref{t:res:perf} shows the most important statistics of our test runs broken down by the dataset groups. To summarize the rows, in total we tried 14,859,756 different instances, of which 6,849,350 (46.1\%) were positive instances, 8,009,719 (53.9\%) were negative instances and 687 (<0.01\%) exceeded the runtime of 10 minutes and therefore timed out. Of the positive instances, 6,478,445 (43.6\%) were immediately solved by Suurballes algorithm for disjoint paths of minimum total length. The remaining 370,905 positive instances (2.5\%) required our search-tree algorithm to yield a solution. Of the negative instances, 7,994,172 (53.8\%) were recognized by the minimum separator heuristic resulting in a separator smaller than $k$. Another 8,090 (0.1\%) were recognized by Suurballes algorithm giving a minimum solution having total length longer than $k\cdot \ell$ and finally 7,457 (0.1\%) required a full traversal of our search tree to determine the infeasibility.

\pagebreak\FloatBarrier 

\begin{figure}[th!]
  \centering
  \includegraphics[width=1.2\textwidth, center]{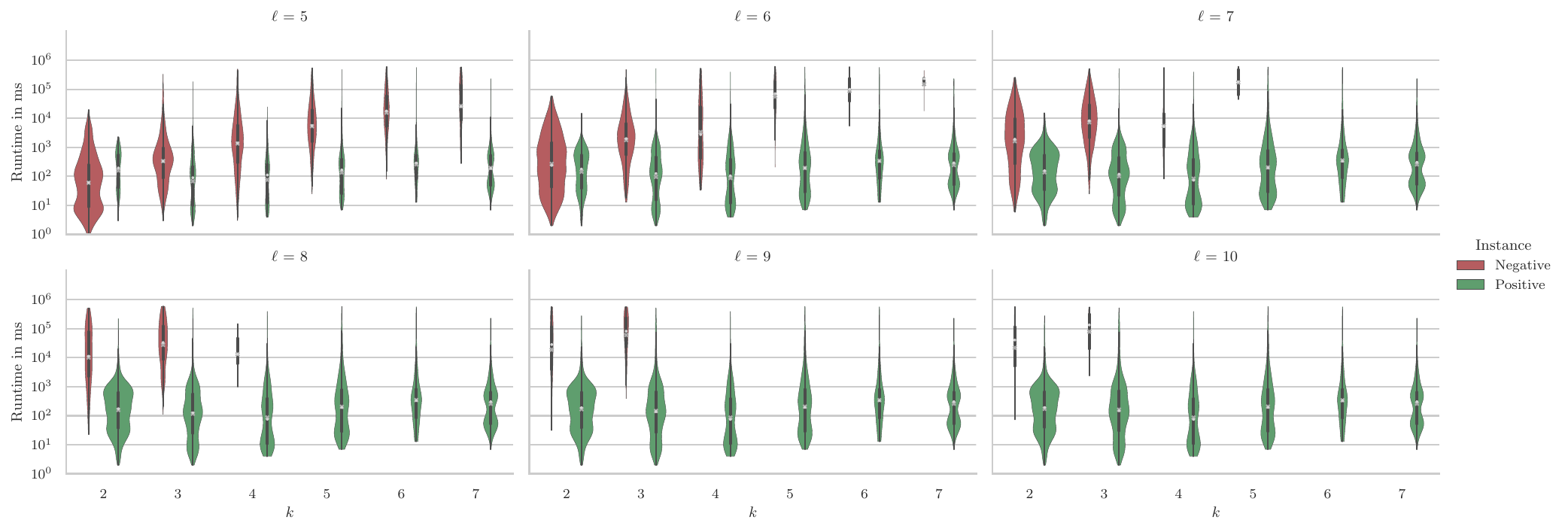}
  \caption{Violin plot showing runtimes for instances from all dataset groups having at least one search tree node, i.e. not being recognized as a trivial instance.}
  \label{fig:res:perf-nopreheur}
\end{figure}

To better analyze the actual runtime of our algorithm, for the remaining observations we discard all fully trivial $s$-$t$-pairs, meaning those pairs for which no combination of $k,\ell$ required more than one search tree node for deciding the instance. However, for all of the pairs that remain we always keep all $k,\ell$ combinations no matter if they individually were trivial. This leaves us with 68,076 instances for \texttt{small}, 107,208 for \texttt{medium} and 74,952 for \texttt{big}. The runtimes of those instances are shown in violin plots in Figure \ref{fig:res:perf}. There, we plot for each dataset group subplots which show the runtime distribution for $\ell \in [5,10]$. Within each subplot, the results are further split by the corresponding values of $k \in [2,7]$ and by whether the instances were positive or negative. The violin plot then models the distribution of the runtimes of positive/negative instances and specific~$k,\ell$~–~the thickness of the violin indicates how many values lie in this range. In all of the plots, instances resulting in a timeout are not displayed, as we know too little about them – neither about their runtime, nor about whether they even are a positive or a negative instance.

\FloatBarrier

Based on those plots, we can make multiple observations: The first obvious effect when looking at the different plots for the three dataset groups is that we can see the overall runtimes growing in Figure \ref{fig:res:perf-medium} and getting even higher in Figure \ref{fig:res:perf-big}. The reason for this most probably is that the polynomial algorithms used within the algorithm take much longer as the graphs get bigger. A second effect can be seen when considering the plots for different values of $\ell$ – for higher values, the number of positive instances is much higher than for small $\ell$. This intuitively makes sense – a higher value of $\ell$ means that we allow more paths to be part of our solution and therefore those instances are more ``relaxed''. 

The most significant observation concerns the runtimes for increasing values of $k$ – there, we can see a clear trend of positive instances taking longer and negative instances getting faster for increasing $k$. This behaviour can also be explained intuitively: For increasing values of $k$, many instances trivially are infeasible, because the minimum separator is smaller than $k$. Therefore, the overall runtimes for negative instances reduce. In this regard, our pre-search-tree heuristics do skew the runtimes. If we want to filter this effect, we can plot only instances requiring at least 1 search tree node. In Figure \ref{fig:res:perf-nopreheur}, we show this for instances of all dataset groups combined. There, we can see that there is also an increase for negative instances with increasing $k$. For higher values of $\ell$, the negative instances exceed the timeout already for values of $k=4$ and therefore are not displayed on the plot anymore.

\subsubsection{Frequency of application of different parts of the algorithm}

\begin{figure}[t]
	\centering
    \begin{subfigure}[b]{.4\textwidth}
         \centering
  \includegraphics[height=15em]{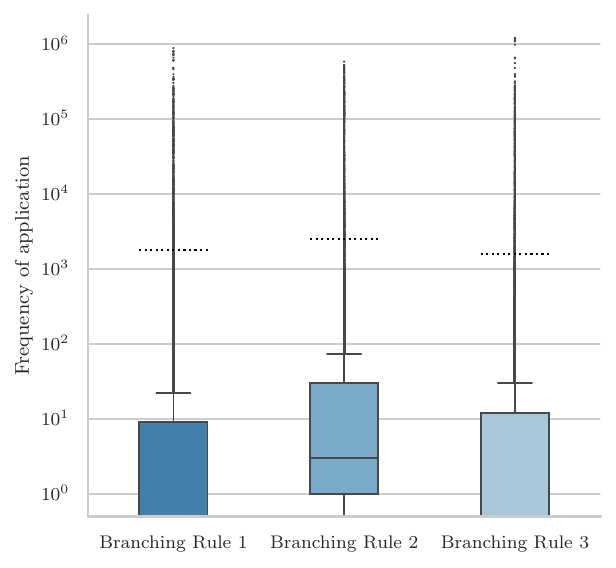}  

  \caption{Frequency of Branching Rules}
             \label{fig:res:stats-br}

     \end{subfigure}
     \hfill
	\begin{subfigure}[b]{.275\textwidth}
         \centering
  \includegraphics[height=15em]{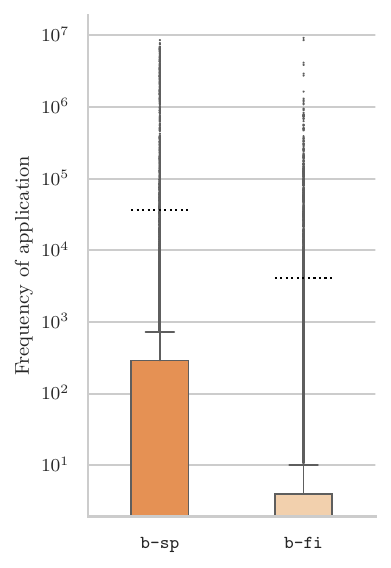}  

  \caption{Frequency of \texttt{b}-Heurs}
             \label{fig:res:stats-bheur}
     \end{subfigure}
	\hfill
    \begin{subfigure}[b]{.275\textwidth}
         \centering
  \includegraphics[height=15em]{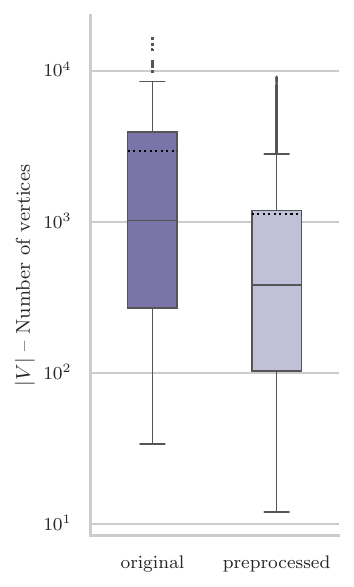}  
  \caption{Effect of Preprocessing}
             \label{fig:res:stats-preproc}

     \end{subfigure}
     	
           \caption{Boxplots showing statistics for the experiments.}
           \label{fig:res:stats}
\end{figure}

Lastly, we want to give insights into the importance of the several mechanisms and improvements employed within the algorithm. For this, we look at the number of times each of the branching rules got invoked (Figure \ref{fig:res:stats-br}), at the frequency of the application of \bsp and \bfi (Figure \ref{fig:res:stats-bheur}) and at the graph sizes before and after preprocessing (Figure \ref{fig:res:stats-preproc}).

We can see, that on average Branching Rule 2, which is invoked by a path in the greedy approach getting too long, is invoked the most. It is followed by Branching Rule 1, which fires when a path in the greedy approach does not exist and closely followed by Branching Rule 3 triggered by the improvement \dms. Even though \bfi has shown great improvements in the previous sections, the main amount of infeasible instances get recognized by the baseline improvement \bsp. Lastly, we can again see that the preprocessing is quite effective, bringing down the average graph size by more than a factor of 3.

\section{Summary of the Results}
\label{ch:exper:dicuss}

Finally, we want to summarize the big picture gained by the experiments. We started by trying running the bare algorithm on our dataset without any improvement, which turned out not to be feasible as the search tree size explodes. Therefore, we settled on a baseline configuration consisting of \bsp, which already gave us promising results. To validate and quantify the effect of this baseline, we additionally tested it on randomly generated small graphs, where also the bare algorithm terminated. This allowed us to compare the performance of \bsp to the bare algorithm on those small graphs, which showed excellent results. 

With this baseline, to understand the effect of the different improvements, we tried each of them individually. The biggest improvements were caused by \bfi, followed by \cdist/\cpl having a strong effect on positive instances, averaging at least at one order of magnitude. In our first tests, \dms had a tendency of worsening the runtimes. However, when applying multiple of the improvements together, in combination with \bfi, \dms seemed to enhance the effect of the forbidden interval rule improving the average runtimes.

It must be noted that while the effects of the optimizations do improve the average runtime, the base configuration \bsp already performed so well that the average improvement was not larger than one order of magnitude. Of course, individually one can find lots of instances where the additional improvements drastically improve the runtime. Nevertheless, it is still interesting that the greatest effect comes from the simplest improvement.

We also investigated the effect of the preprocessing of the graphs, which turned out to be significant, as it allows to speed up all of the polynomial-time algorithms used within our search-tree approach by an average of 50 percent. Another crucial part are the initial heuristics which showed to have high importance as the majority of instances in our test runs were recognized as trivial instances by one of those so-called ``pre''-search-tree heuristics. We showed that without them, especially for negative instances we would have much worse runtimes.

Lastly, concerning the general performance of the algorithm we believe to have achieved quite good results. We found interesting effects on the runtime trends for increasing $k$ and $\ell$, some being counter-intuitive to our presumed worst-case runtime. For example for increasing $k$, the minimum separator heuristic more and more allows us to recognize a negative instance immediately, which is why the average runtimes for those cases even get smaller. Excluding those trivial cases, we can see that the general trend for increasing $k$ and increasing $\ell$ is indeed an exponential growth of the runtimes, as expected. As a last note, it has to be said that all our results are limited within the spectrum of our experiments. Therefore, we can only claim to ``believe'' of having achieved quite good results, as no other reference implementation exists that our approach could have been easily compared to.

%% file: chapters/chapter-conclusion.tex
\onlyinsubfile{\setcounter{chapter}{6}}
\chapter{Conclusion}
\label{ch:concl}

In this thesis, we engineered a new FPT search-tree algorithm for the \SPP problem (SPP), which given a graph $G$, integers $k$ and $\ell$ and vertices $s$~and~$t$, asks for $k$ internally vertex-disjoint $s$-$t$ paths each having a length of at most $\ell$. To our knowledge, only two works on exact algorithms for this problem exist, and none of them seemed to have practical applicability in mind. This is the gap that we wanted to fill with our approach, which is based on a search-tree technique employing greedy localization for bounding the number of subproblems to branch over.

We started by gradually introducing the main branching concept used, which however our initial problem formulation was not able to formalize correctly. Therefore, we introduced a new problem, \SPP \textsc{with Checkpoints} (SPPC) to fully state our search-tree algorithm. The practicability of the algorithm however came by the inclusion of several heuristic improvements which make use of the search tree structure. Relevant improvements included initial techniques for recognizing trivial instances, preprocessing the graph to reduce its size, defining better conditions for recognizing negative instances, intelligently choosing which subproblem to expand next and also improving the greedy localization to further bring down the number of nodes.

The algorithm together with the proposed improvements was implemented and heavily tested against a large set of instances, which showed promising results – on average, most of the randomly chosen instances were solved in under a second. This is highly due to the initial heuristics recognizing trivial instances reliably. For the nontrivial instances, we have witnessed the simplest technique determining shortest paths between checkpoints producing the largest improvement in runtime. But also the other optimizations succeeded in reducing the average runtime noticeably by one order of magnitude.

However, a big weakness of our results is the lack of comparability. Unfortunately, we could not find any existing implementation that we could benchmark our algorithm against. To set our experimental results in relation, a comparison to an ILP implementation would be useful.

Further research could go in two possible directions: On one hand, there is surely more to explore on the theoretical side of the algorithm, meaning e.g. other, more efficient ways of traversing the solution space or even better techniques for shrinking the search tree size. Specifically, the search tree size could be further decreased by investigating more sophisticated preprocessing methods: In the best case, one could be able to determine the exact set of vertices that are part of any $s$-$t$ path of length at most $\ell$. Furthermore, entire biconnected components in the graph could be contracted, as they can only be used in one of the paths. It is also an open question whether we can find a single-exponential search-tree algorithm having a runtime complexity of $2^{O(k\cdot \ell)} \cdot n^{O(1)}$, matching the worst-case complexity of the color coding approach by Golovach and Thilikos. Another perspective on the problem can be given by thinking about possible reduction rules for producing problem kernels.

On the other hand, the implementation of the existing theoretical foundation could be further improved and perfected. It must be said that optimizing the algorithm to get the best possible runtime was never really the focus of this thesis – we rather focused on reducing the search tree size, as this is the dominant factor for the runtime. Any more efficient implementation of the same approach should only differ by polynomial factors in terms of the runtime. However, if one would aim to optimize the runtime, a lot could still be done: One could think of implementing dynamic (all-pair) shortest path algorithms that can efficiently handle small changes caused by temporarily removing some vertices from the graph, because a huge runtime overhead comes from repeatedly computing shortest paths in minor variants of our input graph. 

Another option would be to consider storing computed paths and passing them to child instances, such that when branching, only the paths that have to change can be recomputed. It is not immediately clear whether this would speed up the approach or if it would restrict the instances too much, making it harder to escape local optima. Furthermore, the quality of the different improvements could be analyzed and improvements could be enabled or disabled based on the features of the given instance.

To understand our search-tree algorithm in more depth, the experimental evaluation could be conducted from another perspective, using a random graph model and considering how different levels of sparsity in the graph affect the performance of the algorithm. In this context, the performance of the algorithm on dense graphs for higher values of $k$ and $\ell$ could also be tested, as it is not immediately clear, how hard the problem gets under such conditions.

Furthermore, it would be interesting to investigate whether our search-tree algorithm can be extended to work with other types of graphs like directed graphs or hypergraphs, or when using real-valued edge weights. Similarly, one could try to extend our checkpoint approach to not force the checkpoint lists to start in $s$ and end in $t$, effectively resulting in a multi-terminal variant of \SPp. Additionally, the checkpoints could not only be seen as a tool for implementing the branching strategy, but also considered as a given part of the input, thus implementing a kind of waypoint routing.

Lastly, another interesting possibility would be to investigate certain graph classes for properties which could be exploited to improve the algorithm for those classes and would produce FPT algorithms that might give better bounds. Parameters like bounded treewidth could allow for even better modified algorithms yielding FPT results for those parameters.